
\input phyzzx
\REF\gko     {
P.Goddard, A.Kent and D.Olive \journal Comm.Math.Phys.&103(86)105.
}
\REF\bkr     {
J.Bagger, D.Nemeschansky and S.Yankielowicz \journal Phys.Rev.Lett.&60(88)389;
\hfill\break
D.Kastor, E.Martinec and Z.Qiu \journal Phys.Lett.&200B(88)434;\hfill\break
F.Ravanini \journal Mod.Phys.Lett.&A3(88)397.
}
\REF\bpz    {
A.A.Belavin, A.M.Polyakov and A.B.Zamolodchikov \journal Nucl.Phys.&B241(84)
333.}
\REF\fqs{
H.Eichenherr \journal Phys.Lett.&151B(85)26;
M.A.Bershadskii, V.G.Knizhnik and \hfill\break
M.G.Teitelman, ibid., 31;
D.Friedan, Z.Qiu and S.Shenker, ibid., 97.
}
\REF\zamo   {
A.B.Zamolodchikov and V.A.Fateev \journal Theor.Math.Phys.&71(87)451.
}
\REF\abf   {
G.E.Andrews, R.J.Baxter and P.J.Forrester \journal J.Stat.Phys.&35(84)193.
}
\REF\djkmo   {
E.Date, M.Jimbo, A.Kuniba, T.Miwa and M.Okado \journal Nucl.Phys.&B290(87)231
}
\REF\fefu            {
B.L.Feigin and D.B.Fuchs,in
Representation of Infinite Dimensin Lie Groups,
ed.by A.M.Vershik and D.M.Zhelobenko (Harwood Academic, Chur,
Switzerland, 1988).
}
\REF\eguyan   {
T.Eguchi and S.-K.Yang \journal Mod.Phys.Lett.&A5(90)1693;\hfill\break
T.Eguchi, S.Hosono and S.-K.Yang \journal Comm.Math.Phys.&140(91)159
}
\REF\witten   {
E.Witten \journal Nucl.Phys.&B340(90)281.
}
\REF\dvv   {
R.Dijkgraaf, H.Verlinde and E.Verlinde \journal Nucl.Phys.&B352(91)59.
}
\REF\vafa   {
C.Vafa \journal Mod.Phys.Lett.&A6(91)337.
}
\REF\morozov {
A.Gerasimov, A.Marshakov and A.Morozov
\journal Nucl.Phys.&B328(89)664.
}
\REF\para{
A.B.Zamolodchikov and V.A.Fateev
\journal Sov.Phys.JETP&62(85)215.
}
\REF\felder{
G.Felder \journal Nucl.Phys.&B317(89)215; Erratum
\journal &B324(89)548.
}
\REF\tye{
S.-W.Chung, E.Lyman and S.-H.Tye \journal
Int.J.Mod.Phys.&A7(92)3339.
}
\REF\ocsv{
L.Caneschi, A.Schwimmer and G.Veneziano \journal Phys.Lett.&30B(69)351.
}
\REF\divecchia{
P.DiVecchia, F.Pezzella, M.Frau,
 K.Hornfeck, A.Lerda and S.Sciuto \journal\hfill\break
Nucl.Phys.&B322(89)317
and references there in.
}
\REF\fesi            {
 G.Felder and R.Silvotti
           \journal  Phys.Lett.&231B(89)411
 \journal Comm.Math.Phys.&144(92)17.
}
\REF\alvarez{
C.Vafa \journal Phys.Lett.&190B(86)47;
N.Ishibashi, Y.Matsuo and H.Ooguri \journal\hfil\break
 Mod.Phys.Lett.&A2(87)119;
L.Alvarez-Gaume, C.Gomez, G.Moore
 and C.Vafa \journal Nucl.Phys.&B303(88)455;
N.Kawamoto, Y.Namikawa,
A.Tsuchiya and Y.Yamada \journal Comm.Math.Phys.&116(88)247.
}
\REF\konnow{
H.Konno \journal Phys.Rev.&D45(92)4555.
}
\REF\wakimoto{
M.Wakimoto \journal Comm.Math.Phys.&101(86)605.
}
\REF\geras{
A.Gerasimov, A.Marshakov, A.Morozov, M.Olshanetsky and \hfill\break
S.Shatashvili
\journal Int.J.Mod.Phys.&A5(90)2495.
}
\REF\fms{
D.Friedan, E.Martinec and S.Schenker
\journal Nucl.Phys.&B271
(86)93.
}
\REF\gepwit{
D.Gepner and E.Witten
\journal Nucl.Phys.&B278(86)493.
}
\REF\befe            {
 D.Bernard and G.Felder
           \journal Comm.Math.Phys.&127(90)145.
}
\REF\distqiu         {
J.Distler and Z.Qiu
           \journal Nucl.Phys.&B336(90)533.
}
\REF\kato         {
M.Kato and S.Matsuda
           \journal Adv.Studies in Pure Math.&16(88)205.
}
\REF\frau           {
 M.Frau, A.Lerda, J.G.McCarthy, J.Sidenius
 and S.Sciuto
            \journal Phys.Lett.&245B\hfil\break
            (90)453\journal &254B(91)381.
}
\REF\jns             {
T.Jayaraman, K.S.Narain and M.H.Sarmadi
           \journal Nucl.Phys.&B343(90)418.
}
\REF\bouk          {
P.Bouwknegt, J.G.McCarthy and K.Pilchi
            \journal Nucl.Phys.&B352(91)139.
}
\REF\keli  {
K.Li \journal Nucl.Phys.&B354(91)711.
}
\REF\konnob{
H.Konno \journal Int.J.Mod.Phys.&A7(92)4415.
}
\REF\kape        {
V.G.Ka$\check c$ and D.H.Peterson
           \journal Adv.in.Math.&53 (84)125.
}
\REF\kawai{
T.Kawai, T Uchino and S.-K.Yang, preprint KEK-TH-336.
}
\REF\yoshii{
H.Yoshii\journal Phys.Lett.&B275(92)70.
}
\REF\lerda{
P.Fre, L.Girardello, A.Lerda and P.Soriani, preprint ITP-SB-92-7.
}
\REF\distler{
J.Distler \journal Nucl.Phys.&B342(90)523.
}
\pubnum={RIMS-908}
\date={\quad }
\pubtype={\quad}
\titlepage
\title{
$SU(2)_k\times SU(2)_l/SU(2)_{k+l}$ Coset Conformal Field Theory\break
and Topological Minimal Model on Higher Genus Riemann Surface
}
\author{HITOSHI KONNO}
\address{
Research Institute for Mathematical Sciences\break
Kyoto University, Kyoto 606-01, Japan \break
}
\abstract{
We consider the Feigin-Fuchs-Felder formalism of the
$SU(2)_k\times SU(2)_l/SU(2)_{k+l}$ coset minimal
conformal field theory and extend it to higher genus.
We investigate a double BRST complex with respect to two compatible BRST
charges, one associated with the parafermion sector and
the other associated with  the minimal sector
in the theory.
The usual screened vertex operator is extended to
the BRST invariant screened three string vertex.
We carry out a sewing operation of these string vertices
and derive the BRST invariant screened
$g$-loop operator.
The latter operator characterizes the higher genus
structure of the theory.
 An analogous
operator formalism for the
topological minimal model is obtained as the limit $ l=0$
of the coset theory.
We give
some calculations of  correlation functions on higher genus.
\endpage

\chapter{ Introduction}
      The $SU(2)_k\times SU(2)_l/SU(2)_{k+l}$ minimal coset theory
\refmark{\gko,\bkr} is known
as a rational conformal field theory, which contains a series of minimal
conformal models. The cases $k=1, 2, 4, .. $ provide
the Belavin-Polyakov-Zamolodchikov (BPZ) minimal model,
\refmark{\bpz} the N=1
super minimal model,\refmark{\fqs}
 the $S_3$ symmetric minimal model,\refmark{\zamo}
 and so on.
\par
      It is also a remarkable fact that the theory describes the critical
behavior of the series of
exactly solvable restricted solid on solid (RSOS)
models.\refmark{\abf,\djkmo}
The underlying mathematical structures which connect the RSOS models
and the $SU(2)_k\times SU(2)_l/SU(2)_{k+l}$ coset theory are, however,
unknown. It is, however,  suggestive that
the one point functions in the RSOS models are  given by the
branching coefficients in the coset theory. In other words,
there is a correspondence between the reduced one dimensional
configuration sum in the corner transfer matrix method and the
trace over the irreducible highest weight representation (IHWR)
space in the coset theory.
One of the aim of this paper is to clarify the structure of
IHWR  of the $SU(2)_k\times SU(2)_l/SU(2)_{k+l}$ minimal coset theory
in the Feigin-Fuchs construction.\refmark{\fefu}
\par
      On the other hand, it is known that the
$SU(2)_k\times SU(2)_l/SU(2)_{k+l}$ coset theory
is deeply related to the topological
conformal minimal model. The latter theory was  first obtained by
Eguchi and Yang,\refmark{\eguyan}
 by twisting the N=2 super conformal field theory.
As they showed, the same theory is also realized  as the case
$l=0$ in the $SU(2)_k\times SU(2)_l/SU(2)_{k+l}$ minimal coset theory.
Calculation of correlation functions in the
topological minimal model
has been discussed by Witten\refmark{\witten} and extended by
Dijkgraaf and Verlinde at the tree level,\refmark{\dvv} and
by Vafa to an  arbitrary genus.\refmark{\vafa}
However, since their method totally depends on the correspondence of the
chiral ring structure of the topological conformal field theory
to the one of the super potential in the
topological Landau-Ginzburg theory, its relation to the
Feigin-Fuchs construction, a standard method
in rational conformal field theories,
is unclear.
Our second aim is
to formulate the topological field theory
in the Feigin-Fuchs scheme on Riemann surfaces with arbitrary genus.
\par
      In the next section, we begin
with  reviewing the Feigin-Fuchs construction
of the\break
$SU(2)_k\times SU(2)_l/SU(2)_{k+l}$ minimal coset theory.\refmark{
\bkr,\morozov,\tye}
The theory is represented by the $Z_k$ parafermion theory\refmark{
\para}and
a scalar boson field.
There exist two types of screening charges, one of  which is
just the one in the $Z_k$ parafermion theory and the other
is associated mainly
with the minimal sector in the theory.
\par
      In section 3,
we discuss the  BRST formalism of the theory according to
Felder.\refmark{\felder}
 We treat a double BRST complex associated with the above two
types of screening charges.
We construct the singular vectors explicitly and obtain
the BRST cohomology groups.
\par
      In section 4. we discuss an extension of
the above BRST formalism by introducing the Caneschi-Schwimmer-Veneziano
(CSV) vertex,\refmark{\ocsv}
 a kind of vertex describing the
interaction of three strings.
The CSV vertex connects three different Fock spaces.
Therefore, one can make
a multi loop calculation by taking traces and sewing
of a set of these vertices.\refmark{\divecchia}
One  point to be clarified in  this extension is
a definition of contours for the screening operators.
The contour should be
a homologically nontrivial cycle on the three punctured sphere.
We construct a contour explicitly according to the general method
discussed by Felder and Silvotti.\refmark{\fesi}
We then define a screened CSV vertex.
The BRST relations on  the usual vertex operator
 is hence  extended to the one on the screened CSV vertex operator.
\par
      In section 5, we calculate a BRST invariant $g$-loop operator in the
coset theory. A $g$-loop operator is, in general, used to define a conformal
field theory on a genus $g$ Riemann surface.\refmark{\alvarez}
This is possible, because
a $g$-loop operator
 maps a conformal field to the one on the corresponding
genus $g$ Riemann surface by the Bogoliubov transformation.
Various  Ward identities on a genus $g$ Riemann surface can also be
derived by calculating the action of the currents on the $g$-loop operator.
\par
      Our $g$-loop operator is a generalization of those in the
nondegenerate conformal field theories
in the following two points.\refmark{\konnow}
 The first point is that, in a process of
making loops (i.e. handles),  traces
are taken over the subspace of the Fock space corresponding to
IHWR.
This is,  of course, necessary
to guarantee the BRST invariance, or in other words, the
decoupling of the null states from the $g$-loop operator.
The second point is
the fact that  our $g$-loop operator is properly screened.
This is carried out by
sewing the above screened CSV vertices.
Our screened $g$-loop operator hence provides a consistent
higher genus extension of the Feigin-Fuchs-Felder formalism.
\par
      In section 6, we demonstrate some higher genus calculations
in the coset theory. We give a vacuum amplitude of arbitrary
genus.\par
      The final section is devoted to a discussion of the
topological limit of the coset theory.
Using the results in the coset theory, we formulate
a similar operator formalism
in the topological minimal model. We also give
a possible picture changing operator, which makes the
evaluation of correlators easy.
\par

\chapter{Free Field Realization}
      In this section we  review a free field realization of the
$SU(2)_k\times SU(2)_l/SU(2)_{k+l}$ minimal coset theory.
\refmark{\bkr,\morozov,\tye}
       Let us begin with the  Wakimoto
construction of the $su(2)_k$
affine Ka$\check c$-Moody algebra.
\refmark{\wakimoto}
In the bosonized form, the currents are given by\refmark{\geras}
$$\eqalignno{
J_+&=-e^{\phi}\partial e^{\chi},\cr
J_0&=\partial\phi-\sqrt{k+2\over 2}\partial \Phi,
\cr
J_-&=\Bigl[(k+2)\partial\phi-(k+1)\partial\chi
-\sqrt{2(k+2)}\partial\Phi\Bigr]e^{\phi}e^{-\chi}
,&\eqname{\currents}
\cr}$$
where
$\phi, \chi$ and $\Phi$ are scalar bosons.
These fields satisfy the operator
product expansion (OPE)
$< \chi(z) \chi(w)>
=< \Phi(z)\Phi(w)>=\ln( z-w )=-< \phi(z) \phi(w)>$.
Note that, in \currents, spin 1 conjugate bosons $(\beta, \gamma)$
in the usual Wakimoto
construction are
bosonized as $\beta=-e^{-\phi}\partial e^{\chi}$
and $\gamma=e^{\phi}e^{-\chi}$.\refmark{\fms}
\par
The energy-momentum (EM) tensor defined by the Sugawara form is
obtained as\refmark{\geras}
$$\eqalignno{
T^{(k)}&={1\over 2}(\partial \chi)^2+{1\over 2}\partial^2 \chi
-{1\over 2}(\partial \phi)^2-{1\over 2}\partial^2 \phi
+{1\over 2}(\partial\Phi)^2+{1\over 2}\sqrt{2\over k+2}\partial^2 \Phi.
&\eqname{\emtensor}\cr
}$$
This  generates
the Virasoro algebra with central charge $c^{(k)}={3k\over k+2}$.
The semidirect product of this Virasoro and
the $su(2)_k$ affine Kac-Moody algebra
generates a symmetry in the $SU(2)_k$ Wess-Zumino-Witten (WZW)
 theory.\refmark{\gepwit}
\par
      The primary field $\Phi_{j,j}(z)$ in the $SU(2)_k$ WZW theory,
whose conformal dimension is $h_j={j(j+2)\over k+2}$, is given by
$\Phi_{j,j}=\Phi_{j,m}\vert_{j=m}$, where
$$\eqalignno{
\Phi_{j,m}&=e^{(j-m)(\phi-\chi)}
e^{-j\sqrt{2\over k+2}\Phi}.&\eqname{\primaryf}
\cr}$$
In the unitary representation, the spin $j$ is characterized by an integer $n$
as $2j+1=n$, $1\leq n\leq k+1$. In the following paragraph,
we consider the unitary representation only.
\par
        The screening operator in the $SU(2)_k$ WZW theory is given by
$$\eqalignno{
S^{(k)}&=e^{-\phi}e^{\sqrt{2\over k+2}\Phi}\partial e^{\chi}
.&\eqname{\screenk }\cr}$$
Since the conformal dimension of $S^{(k)}$ is one and its
OPEs with
all the currents in \currents\ yield only
regular or total derivative terms, the
screening charge $\oint S^{(k)}$ is commutable with all the currents.
\par
      Now, let us consider the $SU(2)_k\times SU(2)_l/SU(2)_{k+l}$
coset theory. Let $(\phi^{(a)}, \chi^{(a)}, \Phi^{(a)})$ $a=k, l$ be
the two sets of
the fields realizing $SU(2)_a$ WZW model $a=k,l $ in the way as
\currents.
Let us define new sets of fields
$(\varphi, \chi, \phi_0)$ and
$(\phi^H, \chi^H, \Phi^H)$ by
$$\eqalignno{
\phi^H&=\phi^{(l)}, \qquad \chi^H=\chi^{(l)},\qquad
\Phi^H=
\sqrt{k+2\over p'}\Phi^{(k)}+
\sqrt{p\over p'}\Phi^{(l)}-
\sqrt{2\over p'}\phi^{(k)},\cr
\varphi&=\sqrt{k+2\over k}\phi^{(k)}-\sqrt{2\over k}\Phi^{(k)},\qquad\qquad
\chi=\chi^{(k)},&\eqname{\cosetf}\cr
\phi_0&=
\sqrt{2(k+2)\over kp'}\phi^{(k)}-\sqrt{(k+2)p\over kp'}\Phi^{(k)}+
\sqrt{k\over p'}
\Phi^{(l)},
\cr}$$
where $p=l+2$ and $p'=k+l+2$.
These new fields satisfy
$< \chi^H(z) \chi^H(w)>
=$\break
$< \Phi^H(z)\Phi^H(w)>=
< \chi(z) \chi(w)>
=< \phi_0(z)\phi_0(w)>=\ln( z-w )=-< \phi^H(z) \phi^H(w)>=
-< \varphi(z) \varphi(w)>$. The other OPEs vanish.
Then, the coset theory is realized by a set
$(\varphi, \chi, \phi_0)$, whereas the diagonal $SU(2)_{k+l}$ WZW
model is realized by
 $(\phi^H, \chi^H, \Phi^H)$.\refmark{\morozov}
\par
     The change of fields \cosetf\ makes the total EM
tensor $T^{(k)}+T^{(l)}$ in the $SU(2)_k\times SU(2)_l $ WZW theory
split into two parts
$T^{(k)}+T^{(l)}=T^{(k+l)}+T$, where
$$\eqalignno{
T^{(k+l)}=&
{1\over 2}(\partial \chi^H)^2+{1\over 2}\partial^2 \chi^H
-{1\over 2}(\partial \phi^H)^2-{1\over 2}\partial^2 \phi^H
+{1\over 2}(\partial\Phi^H)^2+{1\over 2}\sqrt{2\over p'}\partial^2 \Phi^H,\cr
T=&T_{Z_k}+
{1\over 2}(\partial\phi_0)^2+{1\over 2}
(2\sqrt{2}\alpha_0)\partial^2 \phi_0&\eqname{\emsplit}\cr
}$$
with  $2\alpha_0=\sqrt{k\over pp'}$ and
$T_{Z_k}$ being  the EM tensor in the $Z_k$ parafermion theory:\refmark{
\distqiu}
$$\eqalignno{
T_{Z_k}=&
{1\over 2}(\partial \chi)^2+{1\over 2}\partial^2 \chi
-{1\over 2}(\partial \varphi)^2-{1\over 2}\sqrt{k\over k+2}\partial^2 \varphi.
&\eqname{\empara}
\cr
}$$
The expression of $T$ indicates
 that the coset theory is realized by the $Z_k$ parafermion theory and
the scalar $\phi_0$ boson theory.
The central charge associated with $T$
is given by
$$\eqalignno{
c&={3kl(k+l+4)\over (k+2)(l+2)(k+l+2)}.
&\eqname{\centc}\cr}$$
One should note the relation: $c^{(k)}+c^{(l)}=c^{(k+l)}+c$.\par
      Next let us  consider the
primary fields and the screening operators.
Let the primary field in the $SU(2)_k\times SU(2)_l $
tensor theory be
$\Phi^{(k)}_{j,j}(z)\Phi^{(l)}_{j',j'}(z)$.
By the field redefinitions in \cosetf\ , one can derive
$$\eqalignno{
\Phi^{(k)}_{j,j}(z)\Phi^{(l)}_{\tilde j,\tilde j}(z)
&=\Phi^{(k+l)}_{\tilde j+j,\tilde j+j}(z)\Psi_{\tilde j,j}(z),
&\eqname{\splitprim}\cr
}$$
where
$$\eqalignno{
\Psi_{\tilde j,j}&=\phi_{\tilde j,\tilde j}
:\exp\lbrace (-\sqrt{2}\alpha_{-}\tilde j-2\sqrt{2}\alpha_0j)\phi_0\rbrace:
&\eqname{\cosetprim}\cr
\phi_{j,m}&=\exp\lbrace
-\sqrt{k+2\over k}(m-{k\over k+2}j)\varphi\rbrace
\exp\lbrace-(j-m)\chi\rbrace&\eqname{\paraprim}\cr
}$$
with $\alpha_{-}=-\sqrt{p\over kp'}$.
Since $\Phi^{(k+l)}_{\tilde j+j,\tilde j+j}$ is the primary field in the
$SU(2)_{k+l}$ WZW model,
the field
 $\Psi_{\tilde j,j}$ in \cosetprim\
can be identified with the primary field in the coset theory.
In fact, in the unitary representation,
the spin $\tilde j$ and $ j$ are given by
$2\tilde j+1=\tilde n$ and
$2 j+1= n$, with integers $\tilde n$ and $n$,
$1\leq\tilde n\leq k+1$ and $1\leq n\leq l+1$.
Introducing the constant $\alpha_{+}=\sqrt{p'\over kp}$ satisfying
$\alpha_{+}+\alpha_{-}=2\alpha_0$ and $\alpha_{+}\alpha_{-}=-1/k$,
we obtain
$$\eqalignno{
\Psi_{J;n',n}&\equiv\Psi_{\tilde j,j} =\phi_{{J\over 2},{J\over 2}}
:\exp\sqrt{2}\alpha_{n',n}\phi_0:
&\eqname{\cosetprimb}\cr
}$$
with $\alpha_{n',n}={1-n'\over 2}\alpha_{-}+{1-n\over 2}\alpha_{+}$
and $J\equiv 2\tilde j=n'-n$ mod $2k$, where $n'=n+\tilde n-1$ satisfying
$1\leq n'\leq k+l+1$ so that $0\leq J\leq k$.
This expression of the primary field has been obtained in Ref.[\bkr].
\par
      The Virasoro highest weight states (the primary states) satisfying
$$\eqalignno{
L_0\vert J;n',n>&=h_{J;n',n}\vert J;n',n>,\cr
L_n\vert J;n',n>&=0,  \qquad for \qquad n\geq1&\eqname{\vhws}\cr
}$$
with the highest weights
$$\eqalignno{
h_{J;n'n}&={J(k-J)\over 2k(k+2)}+{(np'-n'p)^2-k^2\over 4kpp'}
&\eqname{\hweight}\cr
}$$
are created by the primary fields $\Psi_{J;n'n}$
on the $SL(2,C)$ invariant vacuum $\vert 0>$ as
$$\eqalignno{
\vert J;n',n>&=\lim_{z\rightarrow 0}\Psi_{J;n',n}(z)\vert 0>
=\vert J>\otimes\lim_{z\rightarrow 0}e^{\sqrt 2 \alpha_{n'n}\phi_0(z)}\vert 0>.
&\eqname{\hws}\cr}$$
Here $\vert J>$ is the parafermion highest weight state given by
$$\eqalignno{
\vert J>&=\lim_{z\rightarrow 0}\phi_{{J\over2},{J\over2}}(z)\vert 0>.
&\eqname{\parahws}\cr}$$
\par
        Let us denote by $F_{J;n',n}$ the Fock spaces
created by the
action of the creation operators of the fields $\varphi, \chi$ and $\phi_0$
on the primary states \hws.
Here the mode expansion of the field  $X(z)$ is defined by
$$\eqalignno{
X(z)&=ix+\alpha_{X,0}\ln z+\sum_{n\geq1}{1\over \sqrt n}(a_{X,n}^{\dagger}z^n-
a_{X,n}z^{-n}),
&\eqname{\field}\cr}$$
and the following canonical commutation relations are imposed.
$$\eqalignno{
[x,\alpha_{X,0}]&=i\epsilon_X,\qquad
[a_{X,n}, a_{X,m}^{\dagger}]=\epsilon_X\delta_{n,m},
&\eqname{\ccr}\cr}$$
where $\epsilon_X=1$ or -1 for $X=\chi, \phi_0$ or $\varphi$,
respectively. \par
      For later convenience (see \S 3), we formally introduce the field
$$
\eqalignno{
\Psi_{J,m_{n',n};n',n}&=
\phi_{{J\over 2},{m_{n',n}\over 2}}
e^{\sqrt2\alpha_{n',n}\phi_0} &\eqname{\mprime}\cr}$$
with $m_{n',n}= n'-n$ mod $2k$ as well as
the state
$$\eqalignno{
\vert J,m_{n',n};n',n>&=\lim_{z\rightarrow 0}
\Psi_{J,m_{n',n};n',n}(z)\vert 0>.
&\eqname{\mprimestate}
\cr}$$
The conformal dimension of the field $\Psi_{J,m;n',n}$ is
formally given by
$h_{J,m_{n',n};n',n}=h_{J,m_{n',n}}+h_{n',n}$,
where $h_{J,m}={J(J+2)\over 4(k+2)}-{m^2\over 4k}$
and $h_{n',n}={(np'-n'p)^2-k^2\over 4kpp'}$.
The following relations are
satisfied.
$$\eqalignno{
a_{X,s}\vert J,m;n',n>&=0  \qquad for\qquad  s\geq 1 ,\cr
\alpha_{\varphi,0}\vert J,m;n',n>&={1\over 2}\sqrt{k\over k+2}({k+2\over k}m-
J)\vert J,m;n',n> ,\cr
\alpha_{\chi,0}\vert J,m;n',n>&=-{1\over 2}(J-m)\vert J,m;n',n>,
&\eqname{\grounds}\cr
\alpha_{\phi_0,0}\vert J,m;n',n>&=\sqrt 2 \alpha_{n',n}
\vert J,m;n',n>.\cr
}$$
\par
       One should also note that
the state $\vert J>$ plays a similar role
to the spin state in the Ramond sector in the
N=1 super minimal model.
In fact, in the case $k=2$, which corresponds to the
N=1 super conformal minimal model,
the conformal dimension $h_J\equiv {J(k-J)\over 2k(k+2)}$
of the parafermion
primary field $\phi_{{J\over 2},{J\over 2}}$
takes
 0 for $J=0$ mod 2, or ${1\over 16}$ for $J=1$ mod 2.
As is well known, the two sets of primary fields of conformal
dimension \hweight\ with these  $h_J$s
define the Neveu-Schwarz and the Ramond sectors,
respectively.\refmark{\fqs}
Similarly, in the case $k=4$, which corresponds to the $S_3$
symmetric model, $h_J$ takes three values, 0 for $J=0$ mod 4,
${1\over 16}$ for $J=1$ mod 2, or ${1\over 12}$ for $J=2$ mod 4.
In this case,
there exist  three sectors of representation labeled by $h_J$.\refmark{\zamo}
In the case of generic $k$, $h_J$ takes
$\Bigl[{k\over 2}\Bigr]+1$
different values as $J$ varies in the range $0\leq J\leq k$,
where $\Bigl[{k\over 2}\Bigr]$ denotes a maximum integer less than
${k\over 2}$.
Hence the representation space, in this case,
has $\Bigl[{k\over 2}\Bigr]+1$ sectors labeled by  $h_J$.
\par
       In the same way,
the screening operator $S^{(k)}(z)S^{(l)}(z)$
in the tensor theory is  factorized into  $S^{(k+l)}(z)S(z)$.
 One thus obtains
the conformal dimension one operator, up to the  identity
operator $e^{2\sqrt2\alpha_0\phi_0}$,
$$\eqalignno{
S&=e^{-\sqrt{k\over k+2}\varphi}\partial e^{\chi}.
&\eqname{\cscreen}\cr}$$
This operator is nothing but the screening operator
in the
$Z_k$ parafermion theory, as well as
in the $SU(2)_k$ WZW theory.\refmark{\befe,\distqiu}
We hence regard this operator as a screening operator in the coset theory.
\par
     As discussed in [\bkr], there exist another set of screening
operators, namely
$$\eqalignno{
S_{+}&=\Psi e^{\sqrt 2\alpha_{+}\phi_0}, \qquad\qquad
S_{-}=\Psi^{\dagger} e^{\sqrt 2\alpha_{-}\phi_0}.&\eqname{\cscreenpm}
\cr}$$
Here
$\Psi$ and $\Psi^{\dagger}$ are the parafermion currents given
by
$$\eqalignno{
\Psi&=-{1\over \sqrt k}e^{-\sqrt{k+2\over k}\varphi}\partial e^{\chi},\cr
\Psi^{\dagger}&={1\over \sqrt k}(\sqrt{k(k+2)}\partial\varphi
-(k+1)\partial\chi)e^{\sqrt{k+2\over k}\varphi}e^{-\chi}.
&\eqname{\paracurrent}\cr
}$$
\par
      The three screening operators $S_{\pm}$ and $S$ satisfy the
following OPEs.
$$\eqalignno{
S_+(z)S_-(w)&=({1\over (z-w)^2}+{1\over z-w}\sqrt 2\alpha_+
\partial\phi_0(w))
e^{2\sqrt 2\alpha_0\phi_0(w)}+RT,
\cr
S_+(z)S(w)&=RT,&\eqname{\opescreen}\cr
S_-(z)S(w)&=-{k+2\over \sqrt k }\partial_w
({1\over z-w}e^{{2\over k+2}\sqrt{k+2\over k}
\varphi(w)}e^{\sqrt 2\alpha_-\phi_0(w)})+RT.
\cr}$$
\par
      By making use of the primary field $\Psi_{J;n',n}$ and these
screening operators $S$ and $S_{\pm}$, we obtain
the following fusion rules:
$$\eqalignno{
\Bigl[\Psi_{J_1;n_1',n_1}\Bigr]
\times\Bigl[\Psi_{J_2;n_2',n_2}\Bigr]
&=\sum_{n_3=\vert n_1-n_2\vert+1}^{min(n_1+n_2-1,2p-1-n_1-n_2)}
\sum_{n_3'=\vert n_1'-n_2'\vert+1}^{min(n_1'+n_2'-1,2p'-1-n_1'-n_2')}
\Bigl[\Psi_{J_1;n_1',n_1}\Bigr]\cr\eqinsert{\eqname{\fusion}}
}$$

\chapter{BRST Cohomology}
      In this section we
give the definition of  the  BRST charges
associated with $S$ and $S_+$
and  investigate their cohomologies.
\par
      For convenience, let us introduce the
Fock space $F_{J,m_{n',n};n',n}$ defined on the primary
state $\vert J,m_{n',n};n',n>$.
The Fock space $F_{J,m;n',n}$ is  the tensor product of the
parafermion sector $F^{PF}_{J,m}$ and the boson sector $F^{\phi_0}_{n',n}$:
$$\eqalignno{
F_{J,m;n',n}&=F^{PF}_{J,m}\otimes F^{\phi_0}_{n',n}.
&\eqname{\msplitfock}\cr}$$
     \par
 Now let us define the BRST charges $Q_{J+1}$ and $Q^+_n$ by
$$\eqalignno{
Q_{J+1}&={1\over J+1}{e^{2i\pi(J+1)/ k+2 }-1\over
e^{2i\pi/ k+2 }-1 } \oint_{\cal
C}\prod_{i=1}^{J+1}du_i\prod_{i=1}^{J+1}S(u_i),&\eqname{\brsta}
\cr
Q^+_n&={1\over n}{e^{2i\pi n/p }-1\over
e^{2i\pi/p }-1 }
\oint_{\cal C}\prod_{i=1}^{n}dv_i\prod_{i=1}^{n}S_+(v_i),&\eqname{\brstb}
\cr}$$
where the integration contours are taken in the same way as in Ref.[\felder].
Note that the  BRST charge $Q_{J+1}$ is the same one as in the
$Z_k$ parafermion theory.\refmark{\befe,\frau}\par
      It is shown that these charges are nilpotent\refmark{
\felder,\tye}
$$\eqalignno{
Q_{J+1}Q_{k+2-(J+1)}=&
Q_{k+2-(J+1)}Q_{J+1}=0,&\eqname{\nilpoa} \cr
Q^+_{n}Q^+_{p-n}=&
Q^+_{p-n}Q^+_{n}=0&\eqname{\nilpotent}
\cr}$$
and from the OPEs in \opescreen,
these are commuting with each other: $[Q_{J+1},Q^+_{n}]=0$.
\par
      The action of $Q_{J+1}$ and $Q_n^+$ on the Fock space
$F_{J,m_{n',n};n',n}$ give rise to
the following double BRST complex.
$$

  \def\mapright#1{\smash{
      \mathop{\longrightarrow}\limits^{#1}}}
  \def\mapdown#1{\Big\downarrow
      \rlap{$\vcenter{\hbox{$\scriptstyle#1$}}$}}
\eqalignno{
& \matrix{&&\vdots&&\vdots&&\vdots&&\cr
          &&\mapdown{Q_{J+1}}&&\mapdown{Q_{J+1}}&&\mapdown{Q_{J+1}}&&
          \cr
          \cdots&\mapright{Q_n^+}&F_{-J-2,m_{n',-n+2p}}&\mapright{Q^+_{p-n}}&
          F_{-J-2,m_{n',n};n'n}&\mapright{Q^+_n}
          &F_{-J-2,m_{n',-n};n'-n}&\mapright{Q^+_{p-n}}&\cdots\cr
          &&\mapdown{Q_{k+2-(J+1)}}&&\mapdown{Q_{k+2-(J+1)}}&&
          \mapdown{Q_{k+2-(J+1)}}&&
         \cr
         \cdots&\mapright{Q_n^+}&F_{J,m_{n',-n+2p}}&\mapright{Q^+_{p-n}}&
          F_{J,m_{n',n};n'n}&\mapright{Q^+_n}&
          F_{J,m_{n',-n};n',-n}&\mapright{Q^+_{p-n}}&\cdots\cr
          &&\mapdown{Q_{J+1}}&&\mapdown{Q_{J+1}}&&\mapdown{Q_{J+1}}&&\cr
          \cdots&\mapright{Q_n^+}&F_{-J-2,m_{n',-n+2p}}&\mapright{Q^+_{p-n}}&
          F_{-J-2,m_{n',n};n'n}&\mapright{Q^+_n}&
          F_{-J-2,m_{n',-n};n',-n}&\mapright{Q^+_{p-n}}&\cdots\cr
          &&\mapdown{Q_{k+2-(J+1)}}&&\mapdown{Q_{k+2-(J+1)}}&&
          \mapdown{Q_{k+2-(J+1)}}&&\cr
          &&\vdots&&\vdots&&\vdots&&\cr}
&\eqname{\doublcom}
\cr}$$
\par
       Let us first consider the cohomologies of the vertical complexes.
These complexes are  characterized by the
$Z_k$ parafermion sector only. \par
      Since the BRST operator $Q_{J+1}$
in the $Z_k$ parafermion theory is common in the
$SU(2)_k$ WZW theory, the BRST cohomology groups in the parafermion sector
are obtained by the following  theorem.
\subsection{Theorem
(Bernard and Felder)[\befe,\frau]}\par
$$\eqalignno{
Ker Q^{[s]}_{r}/Im Q^{[s-1]}_{r}=
&\Biggl\lbrace\matrix{0&&for\quad s\not=0\cr
                      {\cal H}^{PF}_{J_{r,r'},m}&&for\quad s=0 \cr }
&\eqname{\paracohom}\cr}$$
where ${\cal H}^{PF}_{J_{r,r'},m}$ is the IHWR with highest weight
$h_{J_{r,r'},m}$, $J_{r,r'}=r-r'(k+2)-1$.
In \paracohom, $Q_r^{[2s]}=Q_r$ and $Q_r^{[2s+1]}=Q_{k+2-r}$
act on the parafermion Fock spaces
$
F^{[2s]}_{J_{r,r'},m}=F^{PF}_{J_{r-2s(k+2),r'},m}$ and
$F^{[2s+1]}_{J_{r,0},m}=F^{PF}_{J_{-r-2s(k+2),r'},m}$,
respectively.
As a corollary, we get the trace formula
$$\eqalignno{
\Tr_{{\cal H}^{PF}_{J,m}\otimes F^{\phi_0}_{n,n'}}
{\cal O}&=\sum_{s\in \bf Z}(-)^s
\Tr_{\tilde F^{[s]}_{J,m}\otimes F^{\phi_0}_{n'n}}
 {\cal O}^{\lbrack s\rbrack}, &\eqname{\paratrace}\cr}$$
where ${\cal O}^{\lbrack s\rbrack}$
is an arbitrary operator on $F^{[s]}_{J,m}$, and is obtained
recursively by
$$\eqalignno{
Q^{\lbrack s\rbrack}_{J+1} {\cal O}^{\lbrack s\rbrack}=&
{\cal O}^{\lbrack s+1\rbrack}
Q^{\lbrack s\rbrack}_{J+1} &\eqname{\qxxq}\cr}$$
with ${\cal O}^{\lbrack 0\rbrack}=\cal O$.\par
      In \paratrace,
$\tilde F^{[s]}_{J,m}$ is the Fock space obtained from
$F^{[s]}_{J,m}$ by fixing  picture and dropping redundant zero-modes.
This reduction process is necessary because the Fock space of scalar
boson fields $\phi$ and $\chi$ is
much larger than the one of $\beta$ and $\gamma$
due to  the bosonization.
\par
    In order to find the relation between
 $\tilde F$ and $F$, one has to impose the
constraint $\alpha_{\varphi,0}+\alpha_{\chi,0}=0$
and consider the cohomology of the
nilpotent operator
$ Q_V=\oint dze^{-\chi}$ in the trace.\refmark{\frau,\jns,\bouk}
$Q_V $ commutes with both $Q_{J+1}$ and $Q^+_n$,
and maps the Fock space $F^{[u ]}_{J,m}\equiv
F^{PF}_{J+(k+2)u,m+ku}$ to $F^{[u+1]}_{J,m}$.
 It is easy to show
$$\eqalignno{
Ker  Q^{[u]}_V/Im  Q^{[u-1]}_V&=0,\cr}$$
for all $u\in \bf Z$.
{}From this, we have  the trace
formulae
$$\eqalignno{
\Tr_{\tilde F_{J,m}}{\cal O}&=\sum_{u\geq 0}(-)^u
\Tr_{F^{[u]}_{J,m}}{\cal O}\Bigl\vert_{
\alpha_{\varphi,0}+\alpha_{\chi,0}=0}&\eqname{\vtrace}\cr}$$
and
$$\eqalignno{
0&=\sum_{u\in \bf Z}(-)^u
\Tr_{F^{[u]}_{J,m}}\cal O.
&\eqname{\vtraceb}\cr}$$
Combining \paratrace\ and \vtrace, we get
$$\eqalignno{
\Tr_{{\cal H}^{PF}_{J,m}\otimes F^{\phi_0}_{n,n'}}{\cal O}&
=\sum_{s\in \bf Z}\sum_{u\geq 0}(-)^{s+u}
\Tr_{ F^{[s,u]}_{J,m}\otimes F^{\phi_0}_{n'n}}
{\cal O}^{\lbrack s\rbrack}\Bigl\vert_{
\alpha_{\varphi,0}+\alpha_{\chi,0}=0}. &\eqname{\cparatrace}\cr}$$
\par
       By making use of the resolution in the vertical complexes,
the double complex in
\doublcom\ is now reduced to the following complex
$$\eqalignno{
\cdots \buildrel Q^+_n \over \longrightarrow
{\cal H}^{PF[-1]}_{J,m_{n',n}}\otimes F^{\phi_0[-1]}_{n',n}&
 \buildrel Q^+_{p-n} \over \longrightarrow
{\cal H}^{PF[0]}_{J,m_{n',n}}\otimes F^{\phi_0[0]}_{n',n}
 \buildrel Q^+_{n} \over \longrightarrow
{\cal H}^{PF[1]}_{J,m_{n',n}}\otimes F^{\phi_0[1]}_{n',n}
 \buildrel Q^+_{p-n} \over \longrightarrow \cdots, &
\eqname{\redcom}\cr}$$
where
$$\eqalignno{
{\cal H}^{PF[2t]}_{J,m_{n',n}}\otimes F^{\phi_0[2t]}_{n',n}&
={\cal H}^{PF}_{J,m_{n',n-2tp}}\otimes F^{\phi_0}_{n',n-2tp},\cr
{\cal H}^{PF[2t+1]}_{J,m_{n',n}}\otimes F^{\phi_0[2t+1]}_{n',n}&
={\cal H}^{PF}_{J,m_{n',-n-2tp}}\otimes F^{\phi_0}_{n',-n-2tp}.
&\eqname{\redfock}\cr}$$
\par
       Let us consider
  the singular vectors  in
the Fock space
${\cal H}^{PF}_{J,m_{n',n}}\otimes F^{\phi_0}_{n',n}$.
According to Ref.[\kato]
and the remark given below \grounds,  the singular fields,
in the $h_{J,m}=0$ sector, are given by
$$\eqalignno{
\chi_{-n',n}(z)&=\oint \prod_{i=1}^n d v_i \prod_{i=1}^nS_+(v_i)
e^{\sqrt2\alpha_{-n',n}\phi_0(z)}
&\eqname{\cosnull}\cr}$$
with positive integers $n'$ and $n$. In the generic sector with
nonzero $h_{J,m}$, the singular fields are obtained by
$$\eqalignno{
\chi_{J;-n',n}(z)&=
\chi_{-n',n}(z)\phi_{{J\over2},{m_{-n',n}\over2}}(z)
&\eqname{\cosnullj}\cr}$$
with $J=n'-n$ mod $2k$.
They exist  at the  levels
$$\eqalignno{
h_{J,m_{-n',n};-n',n}-h_{J,m_{n',n};n',n}&=
{4n'n+m^2_{n',n}-m^2_{-n',n}\over 4k}.
&\eqname{\nulllevel}\cr}$$
{}From \cosnull\ and \cosnullj\ ,
we find the following chain of the singular
vectors in
${\cal H}^{PF}_{J,m_{n',n}}\otimes F^{\phi_0}_{n',n}$.
$$\eqalignno{
\matrix{
&\swarrow&\chi_{J;n',-n}&\rightarrow&
\chi_{J;n',n+2p}&
\leftarrow&\chi_{J;n',-n-2p}
&\rightarrow&\chi_{J;n',n+4p}&\leftarrow&\cdots\cr
\Psi_{J,m_{n',n};n',n}&&&{\searrow\swarrow\atop\swarrow\searrow}
&&{\searrow\swarrow\atop\swarrow\searrow}&&
{\searrow\swarrow\atop\swarrow\searrow}
&&{\searrow\swarrow\atop\swarrow\searrow}
&\cr
&\searrow&\chi_{J,n',-n+2p}&\leftarrow&
\chi_{J;n',n-2p}&
\rightarrow&\chi_{J;n',-n+4p}&\leftarrow&\chi_{J;n',n+4p}&\rightarrow&
\cdots\cr}
&\eqname{\coschain}\cr}$$
where an arrow or a chain of arrows from one vector to another means
that the second vector is in the submodule generated by the first vector.
\par
      From \cosnull\ and \cosnullj,
the  primary state $\vert J,m_{n',n};n',n>$ belongs to the
kernel of $Q^+_n$.
In addition, all the singular vectors
in the lower sequence in \coschain\
belong to the kernel of
$Q^+_n$,
because these states are the images of $Q^+_{p-n}$ due to
$$\eqalignno{
\chi_{J;n',n-2ip}&\sim
Q^+_{p-n}\Psi_{J,m_{-n'-(2i-1)p',p-n};-n'-(2i-1)p',p-n},
\cr
\chi_{J;n',-n+2ip}&\sim
Q^+_{p-n}\Psi_{J,m_{n'-(2i-1)p',p-n};n'-(2i-1)p',p-n}
&\eqname{\imnulla}\cr
}$$
for $i\in \bf N$.
We therefore claim that the IHWR
${\cal H}^{G/H}_{J,m_{n',n};n',n}$
with highest weight $h_{J,m_{n',n};n',n}$ in
the coset theory is given by
$$\eqalignno{
Ker Q^+_n/ Im Q^+_{p-n}&={\cal H}^{G/H}_{J,m_{n',n};n',n}.
&\eqname{\cohozero}\cr}$$
\par
       For  the singular vectors  in the Fock spaces
${\cal H}^{PF[t]}_{J,m_{n',n}}\otimes F^{\phi_0[t]}_{n',n}$
with  $t\not=0$,
we find the same  chains as \coschain\ .
For $t<0$, the relations in \imnulla\ as well as the relations
$$\eqalignno{
\chi_{J;n',n+2ip}&\sim Q^+_{n}\Psi_{J,m_{n'-2ip',n};n'-2ip',n},
\cr
\chi_{J;n',-n-2ip}&\sim Q^+_{n}\Psi_{J,m_{-n'-2ip',n};-n'-2ip',n}.
&\eqname{\imnullb}
\cr}$$
for $i\in \bf N$ give rise to the following trivial cohomologies:
$$\eqalignno{
Ker Q_n^{+[t]}/ Im Q_n^{+[t-1]}=&0,
&\eqname{\cohonzero}\cr}$$
where $Q^{+[2t]}_n=Q^+_n$ and $Q^{+[2t+1]}_n=Q^+_{p-n}$.
Furthermore, for $t> 0$,  we have
$$\eqalignno{
\Psi_{J,m_{n',-n-2ip};n',-n-2ip}&\sim Q^+_{n}\chi_{J;n',-n-2ip},
\qquad for \qquad i\geq 0,\cr
\Psi_{J,m_{n',n-2ip};n',n-2ip}&\sim Q^+_{n}\chi_{J;n',n-2(i-1)p},
\qquad for\qquad i\geq 1.&\eqname{\imnullc}
\cr}$$
These relations are derived as follows.
In the Fock space $F_{-J-2,-m_{-n',-n+2ip};-n',-n+2ip}$ dual to
$F_{J,m_{n',n-2ip};n',n-2ip}$, there exists a singular vector
$$\eqalignno{
\chi_{-J-2;-n',n+2ip}&\sim Q^+_n
\Psi_{-J-2,-m_{-n',n+2ip};-n',n+2ip}
&\eqname{\imnulld}\cr}$$
at the level $h_{-J-2,-m_{-n',n+2ip};-n',n+2ip}-
h_{-J-2,-m_{-n',-n+2ip};-n',-n+2ip}$.
Therefore there must be a vector, say  $w_0$,
in
$F_{J,m_{n',n-2ip};n',n-2ip}$
at the same level as
$\chi_{-J-2,-m_{-n',n+2ip};-n',n+2ip}$
such that
$$\eqalignno{
<w_0\vert Q^+_n\Psi_{-J-2,-m_{-n',n+2ip};-n',n+2ip}>&\not=0.
&\eqname{\imnulle}\cr}$$
Since $(Q^+_n)^t=Q^+_n$,
this indicates the existence of a vector $Q^+_n w_0$ ($\not=0$) in \break
$F_{J,m_{n',-n-2ip};n',-n-2ip}$.
This vector is a level zero
vector in $F_{J,m_{n',-n-2ip};n',-n-2ip}$ so that
it can be identified  with the primary
field $\Psi_{J,m_{n',-n-2ip};n',-n-2ip}$.
In addition, since the vector $w_0$ is a level
$h_{J,m_{n',-n-2ip};n',-n-2ip}-h_{J,m_{n',n-2ip};n',n-2ip}$ vector
in $F_{J,m_{n',n-2ip};n',n-2ip}$, $w_0$ is identified with
$\chi_{J;n',-n-2ip}$.
Hence we obtain the first relation in \imnullc\ .
The similar argument implies the second relation.
Combining \imnulla, \imnullb\ and
\imnullc, we claim that the cohomology groups for $t>0$
are trivial, too.\par
      We thus conclude that, for the whole complex in \redcom,
the BRST cohomology groups are given by
$$\eqalignno{
Ker Q_n^{+[t]}/ Im Q_n^{+[t-1]}&
=\Biggl\lbrace\matrix{0&\qquad\qquad for\quad t\not=0\cr
{\cal H}^{G/H}_{J,m_{n',n};n',n}&\qquad\quad for\quad t=0\cr}.
&\eqname{\coscohom}\cr
}$$
As a corollary,  we get a trace formula, which relates the trace over
${\cal H}^{G/H}_{J,m_{n',n};n',n}$ to those over the
Fock spaces ${\cal H}^{PF[t]}_{J,m_{n',n}}\otimes F^{\phi_0[t]}_{n',n}$,
as follows:
$$\eqalignno{
\Tr_{{\cal H}^{G/H}_{J,m_{n',n};n,n'}}{\cal O}&=
\sum_{t\in \bf Z}(-)^{t}
\Tr_{{\cal H}^{PF[t]}_{J,m_{n',n}}
\otimes F^{\phi_0[t]}_{n',n}}
{\cal O}^{\lbrack t\rbrack}
, &\eqname{\costraceb}\cr}$$
where ${\cal O}^{\lbrack t\rbrack}$ are defined recursively by
$$\eqalignno{
Q^{+\lbrack t\rbrack}_n {\cal O}^{\lbrack t\rbrack}=&
{\cal O}^{\lbrack t+1\rbrack}
Q^{+\lbrack t\rbrack}_n &\eqname{\cosrec}\cr}$$
with ${\cal O}^{\lbrack 0\rbrack}={\cal O}$.
\par
      Combining \cparatrace\ and \costraceb, we finally get
$$\eqalignno{
\Tr_{{\cal H}^{G/H}_{J,m_{n',n};n,n'}}{\cal O}&=\sum_{s\in \bf Z}
\sum_{t\in \bf Z}\sum_{u\geq 0}(-)^{s+t+u}
\Tr_{F^{[s,t,u]}_{J,m_{n',n}}\otimes F^{\phi_0[t]}_{n'n}}
{\cal O}^{\lbrack s,t\rbrack}\Bigl\vert_{
\alpha_{\varphi,0}+\alpha_{\chi,0}=0}
.&\eqname{\costrace}\cr}$$
Here
the  Fock space $F^{[s,t,u]}_{J,m_{n',n};n',n}$ is defined by the relations
$$\eqalignno{
F^{[2s,2t,u]}_{J,m_{n',n};n',n}&=F^{[u]}_{J,m_{n',n-2pt};n',n-2pt},\cr
F^{[2s+1,2t,u]}_{J,m_{n',n};n',n}&=F^{[u]}_{-J-2,m_{n',n-2pt};n',n-2pt},
&\eqname{\deffock}\cr
F^{[2s,2t+1,u]}_{J,m_{n',n};n',n}&=F^{[u]}_{J,m_{n',-n-2pt};n',-n-2pt}.\cr
}$$
We also denote an arbitrary physical  operator, which acts
on the Fock space $F^{[s,t,u]}_{J,m_{n',n};n',n}$, by
${\cal O}^{[s,t]}$. Its defining relations are
$$\eqalignno{
{\cal O}^{[s,0]}&={\cal O}^{[s]},&\eqname{\obdef}\cr
Q^{+\lbrack t\rbrack}_n {\cal O}^{\lbrack s,t\rbrack}=&
{\cal O}^{\lbrack s,t+1\rbrack}
Q^{+\lbrack t\rbrack}_n. &\eqname{\obdefb}\cr}$$

\chapter{ Three String Vertex}
    In this section, we discuss the  screened vertex operator and
its  BRST properties.  We next extend it to the screened
three string vertex. This  vertex allows us to extend the
BRST structure discussed in the previous section to the one on
higher genus Riemann surfaces by the  sewing operation.\par
    The screened vertex operator was first introduced by
Felder\refmark{\felder} in the BPZ
minimal model. Its  counterpart in the coset theory is given as
follows
$$\eqalignno{
\Psi_{J;n',n}^{(r,r_+,r_-)}(z)&=\oint_{c} \prod_i^{r}du_i
\oint_{c_+} \prod_j^{r_+}dv_j
\oint_{c_-} \prod_k^{r_-}dw_k
\Psi_{J;n',n}(z) \prod_i^{r}S(u_i) \prod_j^{r_+}S_+(v_j)
\prod_k^{r_-}S_-(w_k),
\cr\eqinsert{\eqname{\scv}}}$$
where the integration contours are taken in the same
way as in Ref.[\felder].
\par
      An important property of the screened vertex
\scv\ is  the following BRST invariance:
$$\eqalignno{
Q_{J+1}\Psi_{A;m',m}^{(r,r_+,r_-)}(z)&=
e^{i\pi (J+1)(-{A\over k+2}-1)}
\Psi_{A;m',m}^{(r,r_+,r_-)}(z)Q_{J+1},
&\eqname{\brstrela}\cr
Q_{k+1-J}\Psi_{A;m',m}^{(r,r_+,r_-)}(z)&=
e^{i\pi (k+1-J)(-{A\over k+2}-1)}
\Psi_{A;m',m}^{(r,r_+,r_-)}(z)Q_{k+1-J},
&\eqname{\brstrelaa}\cr
Q^+_{n}\Psi_{A;m',m}^{(r,r_+,r_-)}(z)&=
e^{i\pi n(-{A\over k}-1+2\alpha_+\alpha_{m',m})}
\Psi_{A;m',m}^{(r,r_+,r_-)}(z)Q^+_{n},
&\eqname{\brstrelb}\cr
Q^+_{p-n}\Psi_{A;m',m}^{(r,r_+,r_-)}(z)&=
e^{i\pi(p-n)(-{A\over k}-1+2\alpha_+\alpha_{m',m})}
\Psi_{A;m',m}^{(r,r_+,r_-)}(z)Q^+_{p-n}.
&\eqname{\brstrelbb}\cr
}$$
\par
      Let us extend this screened vertex to a
screened three string vertex.\refmark{
\fesi,\konnow,\frau}
       There are several  realizations of three string vertex. We here take
the CSV
realization.\refmark{\ocsv,\konnow}
The CSV vertex for the fiels $\varphi, \chi$ and  $\phi_0$
are given by
$$\eqalignno{
<V_{123}\vert&=
<V^{\varphi}_{123}\vert<V^{\chi}_{123}\vert
<V^{\phi_0}_{123}\vert,\cr
<V^X_{123}\vert&=\Bigl[\prod_{r=1,2,3}<\emptyset^r_X\vert\Bigr]
\exp (-\epsilon_X)
\Bigl\lbrace
(\alpha_{X}^{(1)}|\alpha_{X,0}^{(2)})
+(\alpha_{X}^{(2)}|\alpha_{X,0}^{(3)})+(\alpha_{X}^{(3)}|\alpha_{X,0}^{(1)})
\cr
&\qquad+{1\over 2} \sum_{r,s=1\atop(r\not=s)}^3
(\alpha^{(r)}_{X}|D(U_rV_s)|\alpha^{(s)}_{X})\Bigr\rbrace
\delta(\sum_{r=1}^3\alpha^{(r)}_{X,0}+Q_X),&\eqname{\csv}\cr
}$$
where
$(\alpha|\alpha_0)=\sum_{n=1}^\infty\alpha_{n}\alpha_0$,
$(\alpha|D|\alpha)=
\sum_{n,m=1}^\infty
\alpha_{n}D_{nm}\alpha_{m}$
and  $<\emptyset^r_{X}\vert=\sum_{n_X^r}<n^r_X;0|$ with
$<n^r_X;0\vert$ satisfying
$$\eqalignno{
<n^r_X;0\vert \alpha_{X,0}^r&=<n^r_X;0\vert n^r_X\cr
<n^r_X;0\vert \alpha_{X,-n}^r&=0, \qquad for \qquad n\geq 1
&\eqname{\gra}\cr
}$$
for $X=\varphi, \chi$ and $\phi_0$.
 The anomalous charge
 $Q_X$ takes values $\sqrt{k\over k+2}, -1$ and
$-2\sqrt2\alpha_0$  for
 $X=\varphi, \chi$ and $\phi_0$, respectively.
The coefficients $D_{nm}$
are the $(0,0)$ representation matrix elements
of the projective group.
\par
      By using the CSV vertex as a bare vertex,
we define the
screened three string vertex as follows.
$$\eqalignno{
<V^{(r,r_+,r_-)}_{123}>&=\oint_{\cal C}
\prod_{i=1}^{r}du_i
\oint_{\cal C_+}\prod_{j=1}^{r_+}dv_j
\oint_{\cal C_-}\prod_{k=1}^{r_-}dw_k
<V_{123'}||R_{3'3}>\cr
&\qquad\qquad\qquad\times \prod_{i}^rS^{(1)}(u_i)\prod_{j}^{r_+}S_+^{(1)}(v_j)
\prod_{k}^{r_-}S_-^{(1)}(w_k),&\eqname{\sccsv}\cr
}$$
where $|R_{3'3}>$ is
the reflection operator.\refmark{\konnow}
In the expression \sccsv, the integration contours
${\cal C}$, ${\cal C}_+ $
and ${\cal C}_-$ should be taken as
the  $r, r_+$ and $r_-$-homology cycles
defined on the three punctured
sphere. These cycles can be derived by following  the
argument given  by Felder and Silvotti.
\refmark{\fesi}
Their expressions are given in Appendix A.
\par
       The delta function constraints in \sccsv\ give rise to
the following constraints on the (eigen values of) momentum operators.
$$\eqalignno{
\alpha^{(3)}_{\phi_0,0}&=\alpha^{(1)}_{\phi_0,0}+\alpha^{(2)}_{\phi_0,0}+
\alpha_+r_++\alpha_-r_-,\cr
\alpha^{(3)}_{\varphi,0}&=\alpha^{(1)}_{\varphi,0}
+\alpha^{(2)}_{\varphi,0}+
\sqrt{k\over k+2}r+\sqrt{k+2\over k}(r_+-r_-),
&\eqname{\chargecon}\cr
\alpha^{(3)}_{\chi,0}&=\alpha^{(1)}_{\chi,0}+\alpha^{(2)}_{\chi,0}+
r+r_+-r_-.\cr
}$$
\par
      Taking account of these constraints and
the overlapping condition for the bare vertices,\refmark{\konnow}
one can derive the BRST relations for
the screened three
string vertex.
$$\eqalignno{
Q_{J+1}^{(3)}<V^{(r,r_+,r_-)}_{123}>
&=<V^{(r,r_+,r_-)}_{123}>
Q^{(1)}_{J+1}
e^{-\pi i (J+1) (\sqrt{k\over k+2}\alpha^{(2)}_{\varphi,0}
-\alpha^{(2)}_{\chi,0})}\cr
&\qquad\qquad\qquad\qquad+e^{i\pi(A-J)(A+1)/k+2}
<V^{(r+J-A,r_+,r_-)}_{123}>Q^{(2)}_{A+1},
\eqname{\brela}\cr
Q_n^{+(3)}<V^{(r,r_+,r_-)}_{123}>
&=<V^{(r,r_+,r_-)}_{123}>
Q^{+(1)}_n
e^{-\pi i n (\sqrt{k+2\over k}\alpha^{(2)}_{\varphi,0}
-\alpha^{(2)}_{\chi,0}-\sqrt2\alpha_+
\alpha^{(2)}_{\phi_0,0})}\cr
&\qquad\qquad\qquad\qquad+e^{2i\pi(n-m)(n-1)/p}
<V^{(r,r_++n-m,r_-)}_{123}>Q^{+(2)}_m,
\eqname{\brelb}\cr
}$$
where the momentum operator $\alpha^{(3)}_{\varphi,0},
\alpha^{(3)}_{\chi,0}$ and
$\alpha^{(3)}_{\phi_0,0}$
are replaced with their
eigenvalues (see Appendix B).
When the second Fock space in  \brela\ and \brelb\ is saturated
with the highest weight state
$|A;m',m>$,
\brela\ and \brelb\  coincide with the
BRST relations for the screened  vertex operator \brstrela\ and \brstrelb.

\chapter{ Screened Multi-Loop Operators}
    In this section, we derive the screened $g$-loop operator
in  the coset theory.
\par
    Let us first consider the
screened one loop operator.
{}From the BRST relations \brela\ , \brelb\ and
the trace formula \costrace\ ,
the screened one-loop operator
is defined by
$$\eqalignno{
<{\cal\char 84}^{1(r,r_+,r_-)}
_{J,m_{n',n};n',n}\vert&=
{\Tr}^{13}_{{\cal H}^{G/H}_{J,m_{n',n};n',n}}\Bigl\lbrack
<V^{(r,r_+,r_-)}_{123}>\Omega^{(1)\dagger}
q^{L^{(1)}_0-{c\over 24}}
\Omega^{(1)}
\Bigl\rbrack,&\eqname{\scone}
\cr}$$
where $q^{2\pi i\tau}$, $\Omega=e^{L_{-1}}(-)^{L_0}$ and
${\Tr}^{13}$ denotes the trace over the first and the third
Fock spaces.\refmark{\konnow}
\par
     Carrying out the trace ${\Tr}^{13}$ explicitly,
we obtain the expressions for the screened one-loop operator as follows.
$$\eqalignno{
<{\cal T}^{1(r,r_+,r_-)}_{J;n',n}|&=\oint_{\cal C}
\prod_{i=1}^rdu_i
\oint_{\cal C_+}\prod_{j=1}^{r_+}dv_j
\oint_{\cal C_-}\prod_{k=1}^{r_-}dw_k
<{\cal T}_{J;n',n}^{(1)}|
\prod_{i=1}^r
S(u_i)
\prod_{j=1}^{r_+}
S_+(v_j)\prod_{k=1}^{r_-}S_-(w_k)\cr
&\times\delta(
\alpha_{\phi_0,0}+\sqrt2\alpha_+r_++\sqrt2\alpha_-r_-)
\cr
&\times\delta(
\alpha_{\varphi,0}+\sqrt{k\over k+2}r+\sqrt{k+2\over k}(r_+ - r_-))
\cr
&\times\delta(\alpha_{\chi,0}+r+r_+ - r_-),
&\eqname{\sconea}
\cr}$$
where
$$\eqalignno{
<{\cal T}_{J;n',n}^{(1)}|&=
\sum_{s\in \bf Z}
\sum_{t\in \bf Z}
\sum_{u\geq 0}
(-)^{s+t+u}
<{\cal T}_{J}^{(1)\varphi\chi[s,t, u]}|
<{\cal T}_{n',n}^{(1)\phi_0[t]}|
e^{-i\pi\theta^{[s,t]}}
&\eqname{\sconetad}\cr
<{\cal T}_{J}^{(1)\varphi\chi[s,t,u]}|&=
\eta(\tau)^{-2}q^{B^{[s,t,u]}_{\varphi\chi}}
<\emptyset_{\varphi\chi}|\exp\Bigl\lbrace
2\pi i\Bigl[p_{\varphi}^{[s,t]}(\alpha_{\varphi}\vert+
p_{\chi}^{[s,t,u]}(\alpha_{\chi}\vert\Bigr]
|A)\cr
&\qquad\qquad+\Bigl[Q_{\varphi}(\alpha_{\varphi}\vert
+Q_{\chi}(\alpha_{\chi}\vert\Bigr]
|C)
+{1\over 2}\Bigl[(\alpha_{\chi}|Q|\alpha_{\chi})-
(\alpha_{\varphi}|Q|\alpha_{\varphi})\Bigr]\Bigr\rbrace,
&\eqname{\sconeb}
\cr
<{\cal T}^{(1)\phi_0[t]}_{n',n}|
&=\eta(\tau)^{-1}q^{B^{[t]}_{\phi_0}}
<\emptyset_{\phi_0}\vert
\exp\Bigl\lbrace 2\pi ip^{[t]}_{\phi_0}(\alpha_{\phi_0}|A)
+Q_{\phi_0}(\alpha_{\phi_0}|C)
 +{1\over2}(\alpha_{\phi_0}|Q|\alpha_{\phi_0})
\Bigr\rbrace\cr\eqinsert{\eqname{\sconec}}
}$$
with
$\theta^{[2s,2t]}=0$,
$\theta^{[2s+1,2t+1]}=\theta^{[2s+1,2t]}+\theta^{[2s,2t+1]}$
and
$$\eqalignno{
&\theta^{[2s+1,2t]}=(J+1)\Bigl(\sqrt{k\over k+2}\alpha_{\varphi,0}
-\alpha_{\chi,0}\Bigr),&\eqname{\thetaa}\cr
&\theta^{[2s,2t+1]}=n\Bigl(\sqrt{k+2\over k}\alpha_{\varphi,0}
-\alpha_{\chi,0}-\sqrt2\alpha_+\alpha_{\phi_0,0}\Bigr),
&\eqname{\thetb}\cr
&B^{[s,t,u]}_{\varphi\chi}=(k+2)\Bigl(\Bigl[{s\over 2}\Bigr]
-{u\over 2}-{J^{[s]}+1\over 2(k+2)}\Bigr)^2
-k\Bigl({m^{[t]}\over 2k}+{u\over 2}\Bigr)^2,&\eqname{\bvarphi}\cr
&B^{[t]}_{\phi_0}={1\over 4kpp'}\Bigl(n'p-(-)^tnp'
+2pp'\Bigl[{t\over 2}\Bigr]\Bigr)^2,&\eqname{\bphi}\cr
&p^{[s,t]}_{\varphi}=-{1\over 2}\sqrt{k\over k+2}
\Bigl({k+2\over k}m^{[t]}-J^{[s]}
+2(k+2)\Bigl[{s\over 2}\Bigr]\Bigr),&\eqname{\pvarphi}\cr
&p^{[s,t,u]}_{\chi}=-{1\over 2}
\Bigl(J^{[s]}-2(k+2)\Bigl[{s\over 2}\Bigr]-m^{[t]}+2u\Bigr),&\eqname{\pchi}\cr
&p^{[t]}_{\phi_0}=\sqrt2\alpha_{n',(-)^tn-2p[{t\over 2}]}.&\eqname{\pphi}\cr
}$$
We also use the notations  $J^{[2s]}=J, J^{[2s+1]}=-J-1$ and $m^{[t]}=
m_{n',(-)^tn-2p[{t\over 2}]}$.
The coefficients $A, C$ and $Q$ in \sconeb\  and
\sconec\ are given by
$$\eqalignno{
A_n=&{1\over n!}\partial_x^n \int^xd\omega\vert_{x=0},\qquad
&\eqname{\coeffa}\cr
C_n=&{1\over n!}\partial_x^n\ln\sigma(x)\vert_{x=0},\qquad
&\eqname{\coeffc}\cr
Q_{nm}=&{1\over n!m!}
\partial_x^n
\partial_y^m \ln{E(x, y)\over x-y}
\vert_{x=y=0},&\eqname{\coeffq}\cr
}$$
where  $d\omega$,
$\sigma(x)$ and $E(z,w)$
are
the first Abelian differential,
the ${g\over2}$-differential
and the prime form on the genus $g$ Riemann surface
( $g$=1 for the above case ),
respectively.
The expressions for these functions
in terms of
the Schottky variables are given in Ref.\lbrack\divecchia\rbrack.
\par
     From \brela\ and \brelb\ ,
the BRST relation for the screened loop operator
is given by
$$\eqalignno{
<{\cal T}^{1(r,r_+,r_-)}_{J;n',n}\vert Q_{\tilde J+1}&=0,
&\eqname{\binva}\cr
<{\cal T}^{1(r,r_+,r_-)}_{J;n',n}\vert Q^+_{\tilde n}&=0
&\eqname{\binv}\cr}$$
with $\tilde J=r$ and $\tilde n=2r_++1$.
 These relations
indicate the decoupling of the null states from the screened loop
operator.
\par
       The $g$ loop extension is obtained
by sewing the $g$ screened one loop operators \sconea\ by
the $g$-1 screened three string vertices  \sccsv.\refmark{\konnow}
The result is
$$\eqalignno{
<{\cal T}^{g(r,r_+,r_-)}_{\lbrack {\cal J};N',N\rbrack}|&=
\prod_{a=1,\cdots,g\atop
I_1,\cdots,I_{g-1}}
\Bigl[\oint_{{\cal C}^a}
\prod_{i=1}^{r^a}du_i^a
\oint_{{\cal C}^a_+}
\prod_{j=1}^{r^a_+}dv_k^a
\oint_{{\cal C}^a_-}
\prod_{k=1}^{r^a_-}dw_k^a\Bigr]\cr
&\times <{\cal T}^{(g)}_{\lbrack {\cal J};N',N\rbrack}|
\prod_{a=1,\cdots,g\atop
I_1,\cdots,I_{g-1}}\Bigl[\prod_{i=1}^{r^a}S(u_i^a)
\prod_{k=1}^{r^a_+}S_+(v_j^a)
\prod_{k=1}^{r^a_-}S_-(w_k^a)\Bigr]
\cr
&\times\delta(
\alpha_{\phi_0,0}+\sqrt2\alpha_+r_++\sqrt2\alpha_-r_-+2\sqrt2\alpha_0(g-1))
\cr
&\times\delta(
\alpha_{\varphi,0}+\sqrt{k\over k+2}r+\sqrt{k+2\over k}(r_+ - r_-)
+\sqrt{k\over k+2}(g-1))
\cr
&\times\delta(\alpha_{\chi,0}+r+r_+ - r_-),
&\eqname{\scg}
\cr}$$
where $r=\sum_{a=1,\cdots,g,\atop
I_1,\cdots,I_{g-1}}r^a,\quad r_+=\sum_{a=1,\cdots,g,\atop
I_1,\cdots,I_{g-1}}r_+^a,\quad r_-=\sum_{a=1,\cdots,g,\atop
I_1,\cdots,I_{g-1}}r_-^a $ and
$$\eqalignno{
<{\cal T}^{g}_{\lbrack {\cal J};N',N\rbrack}|
&=
\prod_{a=1}^g \Bigl[
\sum_{s_a\in \bf Z}
\sum_{t_a\in \bf Z}
\sum_{u_a\geq 0}
(-)^{s_a+t_a+u_a}\Bigr]
<{\cal T}^{(g)\varphi\chi[S,T,U]}_{\lbrack {\cal J}\rbrack}|
<{\cal T}^{(g)\phi_0[T]}_{{[N',N]}}|
\prod_{a=1}^ge^{-i\pi\theta^{[s_a,t_a]}}
\cr
<{\cal T}^{(g)\varphi\chi[S,T,U]}_{\lbrack {\cal J}\rbrack}|
&=
F[G^{(g)}]^{-2}
<\emptyset_{\varphi\chi}|\exp\Bigl\lbrace
\sum_{X=\varphi,\chi}\Bigl[
\pi i\epsilon_X \sum_{a,b=1}^g
p_{X}^{[s_a,t_a,u_a]}
\tau_{ab}p_{X}^{[s_b,t_b,u_b]}
\cr
&\qquad\qquad\qquad +Q_X(\alpha_{X}|C)
+\epsilon_X{1\over 2}(\alpha_{X}|Q|\alpha_{X})
\cr
&\qquad\qquad\qquad-
2\pi i\sum_{a=1}^g
p_{X}^{[s_a,t_a,u_a]}
[(\alpha_{\Phi}|A^{a})+
Q_X(\triangle^{(g)a}+{1\over 2})]
\Bigr]\Bigr\rbrace,
&\eqname{\scgb}\cr
<{\cal T}^{(g)\phi_0}|
&=
F[G^{(g)}]^{-1}
<\emptyset_{\phi_0}\vert
\exp \Bigl\lbrace \pi i \sum_{a,b=1}^g
p_{\phi_0}^{[t_a]}
\tau_{ab}
p_{\phi_0}^{[t_b]}+Q_{\phi_0}(\alpha_{\phi_0}|C)
+{1\over 2}
(\alpha_{\phi_0}|Q|\alpha_{\phi_0})\cr
&\qquad\qquad\qquad-
2\pi i\sum_{a=1}^g
p_{\phi_0}^{[t_a]}
[(\alpha_{\phi_0}|A^{a})
+Q_{\phi_0}(\triangle^{(g)a}+{1\over 2})]
\Bigr\rbrace
&\eqname{\scgc}
 \cr}$$
with $\theta^{[2s_a,2t_a]}=0$,
 $\theta^{[2s_a+1,2t_a+1]}=\theta^{[2s_a+1,2t_a]}+\theta^{[2s_a,2t_a+1]}$
and
$$\eqalignno{
\theta^{[2s_a+1,2t_a]}&={2(J+1)r^a\over k+2},\qquad
\theta^{[2s_a,2t_a+1]}={2nr_+^a\over p}.&\eqname{\bdefphase}\cr
}$$
In the above, ${\cal J}=\lbrace J_a\rbrace$,
$N^{(')}=\lbrace n^{(')}_a\rbrace$,
$S=\lbrace s_a\rbrace,
T=\lbrace t_a\rbrace$ and $U=\lbrace u_a\rbrace, a=1,2,..g$.
We also denote by $\tau$ the period matrix,
by $\triangle$ the Riemann constant and
by $F[G^{(g)}]$ the partition function associated with
the Schottky group $G^{(g)}$.
$F[G^{(g)}]$ is related to the scalar
determinant as follows.\refmark{\konnob}
$$\eqalignno{
F[G^{(g)}]&=(\det \bar\partial_0)^{1/2}.
&\eqname{\partdet}\cr}$$
The coefficients
$A,C $ and $Q$ are given by \coeffa\ $\sim$ \coeffc\ with using
the functions $\varphi, \sigma$ and $E$ defined on the genus $g$ Riemann
surface.
\par
      Taking account of
 the BRST invariance of the screened CSV vertex \brela\ and \brelb
as well as those for the screened one loop operator \binva\ and
\binv, one can show the BRST invariance relations
for the screened $g$-loop operator analogous
to \binva\ and \binv.\par

\chapter{Conformal Blocks on Higher Genus}
       The loop operator maps the fields on the sphere to the one
on the corresponding higher genus Riemann surface.
This allows one to calculate any higher genus conformal block functions.
\par
       In general, a  genus $g$ conformal block function
can be evaluated by
$$\eqalignno{
<\prod_{i=1}^L\Psi_{ J_i;n_i',n_i}(z_i)>_{g,[{ \cal J};N',N]}
&=<{\cal T}^{g(r,r_+,r_-)}_{[{\cal J},N',N]}\vert
\prod_{i=1}^N\Psi_{ J_i;n_i',n_i}(z_i)\vert 0>,
&\eqname{\confblock}\cr}$$
up to an insertion of certain identity operators (see later paragraphs).
The delta function constraints
on the screened loop operator give rise to  the following
selection rules for this conformal block function.
$$\eqalignno{
&\sum_{i=1}^L\alpha_{n_i',n_i}+\alpha_+r_++\alpha_-r_-+2\alpha_0(g-1)=0,
\cr
&\sum_{i=1}^L{J_i\over \sqrt{ k(k+2)}}+\sqrt{k\over k+2}r+\sqrt{k+2\over k}(r_+
- r_-)
+\sqrt{k\over k+2}(g-1)=0,
&\eqname{\selectr}\cr
&r+r_+ - r_-=0.
\cr}$$
These relations determine $r, r_+$ and $r_-$.
\par
       Let us consider the $g$=1 zero point function, i.e. the character of the
Virasoro
algebra in the coset theory.
By making use of \scone, we obtain
$$\eqalignno{
\chi_{J;n',n}(\tau)&=<{\cal T}^{g(0,0,0)}_{J;n',n}\vert\vert 0>\cr
&=\eta(\tau)^{-3}\sum_{s\in\bf Z}\sum_{t\in\bf Z}\sum_{u\geq 0}(-)^{s+t+u}
q^{B^{[s,t,u]}_{\varphi\chi}+B^{[t]}_{\phi_0}}\cr
&=\sum_{m\in\bf Z}C_{J,m}(\tau)\Bigl( \sum_{t\in\bf Z}\delta_{m,m_{n',n-2pt}}
q^{B_{\phi_0}^{[2t]}}-\sum_{t\in\bf Z}\delta_{m,m_{n',-n-2pt}}
q^{B_{\phi_0}^{[2t+1]}}\Bigr),
&\eqname{\charact}\cr
}$$
where $C_{J,m}$ is the string fuction given by\refmark{\kape}
$$\eqalignno{
C_{J,m}(\tau)&=\sum_{s\in\bf Z}\sum_{u\geq 0}(-)^{u}
\Bigl(q^{B_{\varphi\chi}^{[2s,u]}}-q^{B_{\varphi\chi}^{[2s+1,u]}}\Bigr).
&\eqname{\stringf}\cr}$$
The result \charact\ is nothing but the branching coefficient.\refmark{
\bkr,\tye}
\par
         Let us next extend this result to higher genus.
Let us call a genus $g$ vacuum (zero point) amplitude as a genus $g$ character.
The selection rules for the  genus $g$ character are given by
$$\eqalignno{
&\alpha_+r_++\alpha_-r_-+2\alpha_0(g-1)=0,
\cr
&\sqrt{k\over k+2}r+\sqrt{k+2\over k}(r_+ - r_-)
+\sqrt{k\over k+2}(g-1)=0,&\eqname{\selectg}
\cr
&r+r_+ - r_-=0.
\cr}$$
Solving these equations, one gets
$$\eqalignno{
r&={k\over 2}(g-1),\qquad
r_+={l\over 2}(g-1),\qquad
r_-={k+l\over 2}(g-1).&\eqname{\numbro}\cr
}$$
Since the total number of screening charges must be an integer,
these conditions are satisfied only for $g$ being an odd integer.
The same phenomenon has been discussed
in the BPZ minimal model,\refmark{
\fesi}as well as in the $SU(2)_k$ WZW model.\refmark{\konnow,\frau}
For the enen genus character, one has to insert the identity
operator
$$\eqalignno{
I&=e^{-\sqrt{k\over k+2}\varphi}e^{2\sqrt2\alpha_0\phi_0}.
&\eqname{\idop}
\cr}$$
This shifts the $\phi_0$ and the $\varphi$ charge
so that one gets the new conditions
$$\eqalignno{
r&={k\over 2}g,\qquad
r_+={l\over 2}g,\qquad
r_-={k+l\over 2}g.&\eqname{\numbre}\cr
}$$
Under these conditions, one obtains the expressions for the
genus $g$ characters as follows.
$$\eqalignno{
\chi&^{(g)}_{[{\cal J};N'N]}(\tau)=
<{\cal T}^{g(r,r_+,r_-)}_{[{\cal J},N',N]}\vert I(Q)
 \vert 0>\cr
&=F[G^{(g)}]^{-3}\oint d\Omega
\prod_{a=1}^g \Bigl[
\sum_{s_a\in \bf Z}
\sum_{t_a\in \bf Z}
\sum_{u_a\geq 0}
(-)^{s_a+t_a+u_a}e^{-i\pi\theta^{[s_a,t_a]}}\Bigr]
\triangle^{(g)}_{[{\cal J},N',N;S,T,U ]}(\tau)\cr
&\times{\cal E}^{(g)}(u,v,w)
\sum_{K=K_+\cup K_-}
\prod_{I}{\partial \over\partial \rho_I}
\prod_{J}{\partial \over\partial \lambda_J}
\prod_{K_+}{\partial \over\partial \mu_{K_+}}
\prod_{K_-}{\partial \over\partial \nu_{K_-}}
I^{(g)}(\rho,\lambda,\mu,\nu)\Bigl\vert_0,
&\eqname{\gcharac}\cr}$$
where
 $\lbrace u_I\rbrace_{I=1,2...r}=\lbrace u^1_1,
...,u^g_{r^g},u^{I_1}_1,...,u^{I_{g-1}}_{r^{I_{g-1}}}\rbrace$,
 $\lbrace v_J\rbrace_{J=1,2...r_+}=\lbrace v^1_1,
...,v^g_{r_+^g},v^{I_1}_1,...,v^{I_{g-1}}_{r_+^{I_{g-1}}}\rbrace$
and  $\lbrace w_K\rbrace_{K=1,2...r_-}=\lbrace w^1_1,
...,w^g_{r_-^g},w^{I_1}_1,...,w^{I_{g-1}}_{r_-^{I_{g-1}}}\rbrace$.
In \gcharac,
the sum $\sum_{K=K_+\cup K_-}$ should be taken over all the ways of
decomposition of the index set $K$ into two disjoint sets $K_+$ and $K_-$, and
$I(Q)$ should be inserted only in the even genus case.
We also use the notation
$$\eqalignno{
\oint d\Omega&=
\prod_{a=1,\cdots,g\atop
I_1,\cdots,I_{g-1}}
\Bigl[\oint_{{\cal C}^a}
\prod_{i=1}^{r^a}du_i^a
\int_{{\cal C}^a_{+}}
\prod_{j=1}^{r^a_+}dv_k^a
\oint_{{\cal C}^a_-}
\prod_{k=1}^{r^a_-}dw_k^a\Bigr].
&\eqname{\meas}
\cr}$$
For odd genus, we obtain
$$\eqalignno{
&\triangle^{(g:odd)}_{[{\cal J},N',N;S,T,U ]}(\tau)\crr
&=\exp\Bigl\lbrace
i\pi\sum_{a,b=1}^g\Bigl[
p^{[t_a]}_{\phi_0}\tau_{ab}p^{[t_b]}_{\phi_0}
+p^{[s_a,t_a,u_a]}_{\chi}\tau_{ab}p^{[s_b,t_b,u_b]}_{\chi}
-p^{[s_a,t_a]}_{\varphi}\tau_{ab}p^{[s_b,t_b]}_{\varphi}
\Bigr]
\crr
&\qquad\qquad\qquad+2\pi i\sum_{a=1}^g\Bigr[
\Bigl(p^{[s_a,t_a,u_a]}_{\chi}+\sqrt{k+2\over k}p^{[s_a,t_a]}_{\varphi}\Bigr)
\sum_{I}\int^{u_I}_{P_0}d\omega^a\crr
&\qquad\qquad\qquad
-\Bigl(\sqrt2\alpha_+p^{[t_a]}_{\phi_0}+p^{[s_a,t_a,u_a]}_{\chi}
+\sqrt{k+2\over k}p^{[s_a,t_a]}_{\varphi}
\Bigr)\sum_{J}\int^{v_J}_{P_0}d\omega^a\crr
&\qquad\qquad\qquad
+\Bigl(\sqrt2\alpha_-p^{[t_a]}_{\phi_0}-p^{[s_a,t_a,u_a]}_{\chi}
-\sqrt{k+2\over k}p^{[s_a,t_a]}_{\varphi}
)\sum_{K}\int^{w_K}_{P_0}d\omega^a\crr
&\qquad\qquad\qquad+
\Bigl(-2\sqrt2\alpha_0p^{[t_a]}_{\phi_0}+p^{[s_a,t_a,u_a]}_{\chi}
-\sqrt{k\over k+2}p^{[s_a,t_a]}_{\varphi}
\Bigr)\int^{\triangle}_{(g-1)P_0}d\omega^a
\Bigr]\Bigr\rbrace,&\eqname{\triodd}\cr
&{\cal E}^{(g:odd)}(u,v,w)\crr
&=\Bigl[{\prod_{I,J}E(u_I,v_J)\over
\prod_{I<I'}E(u_I,u_{I'})
\prod_{I,J}E(u_I,v_J)}\Bigr]^{2\over k}
{\prod_{J}\sigma(v_J)^{-{2\over p}}
\prod_{K}\sigma(w_K)^{2\over p'}
\over \prod_{J,J'} E(v_J,v_{J'})^{{2\over k}-1}
\prod_{K<K'} E(w_K,w_{K'})^{{2\over k}-1}
},&\eqname{\eodd}\cr
&I^{(g:odd)}(\rho,\lambda,\mu,\nu)\crr
&=\exp\Bigl\lbrace 2\pi i\sum_{a=1}^g\Bigl[
\sqrt{k+2\over k}p^{[s_a,t_a]}_{\varphi}
\sum_{K_+}\mu_{K_+}\omega^a(w_{K_+})\crr
&\qquad\qquad\qquad\qquad
-p_{\chi}^{[s_a,t_a,u_a]}
[\sum_{I}\rho_{I}\omega^a(u_I)+\sum_{J}\lambda_J\omega^a(v_J)
-\sum_{K_-}\nu_{K_-}\omega^a(w_{K_-})]\Bigr]\Bigr\rbrace\crr
&\qquad
\times\prod_Ie^{-\rho_{I}\partial\ln\sigma(u_I)}\prod_Je^{-\lambda_{J}\partial\ln\sigma(v_J)}
\prod_{K_+}e^{-\mu_{K_+}\partial\ln\sigma(w_{K_+})}\prod_{K_-}
e^{-\nu_{K_-}\partial\ln\sigma(w_{K_-})}\crr
&\qquad\times
\prod_{I,K_+}e^{({k+2\over k}\mu_{K_+}\partial_{w_{K_+}}
-\rho_{I}\partial_{u_{I}})
\ln E(u_{I},w_{K_+})}
\prod_{J,K_+}e^{({k+2\over k}\mu_{K_+}\partial_{w_{K_+}}
-\lambda_{J}\partial_{v_{J}})
\ln E(v_{J},w_{K_+})}\crr
&\qquad\times
\prod_{K_-,K_+}e^{-({k+2\over k}\mu_{K_+}\partial_{w_{K_+}}
+\nu_{K_-}\partial_{w_{K_-}})
\ln E(w_{K_-},w_{K_+})}\crr
&\times
\prod_{I<I'}e^{D(\rho_I,\rho_{I'};u_I,u_{I'})
\ln E(u_I,u_{I'})}
\prod_{J<J'}e^{D(\lambda_J,\lambda_{J'};v_J,v_{J'})
\ln E(v_J,v_{J'})}
\prod_{I,J}e^{D(\rho_I,\lambda_{J};u_I,v_{J})
\ln E(u_I,v_{J})}\crr
&\qquad\times\prod_{I,K_-}e^{D(\rho_{I},\nu_{K_-};u_{I},w_{K_-})
\ln E(u_{I},w_{K_-})}
\prod_{J,K_-}e^{D(\lambda_{J},\nu_{K_-};v_{J},w_{K_-})
\ln E(v_{J},w_{K_-})}\crr
&\times\prod_{K_-<K_-'}e^{D(\nu_{K_-},\nu_{K_-'};w_{k_-},w_{K_-'})
\ln E(w_{K_-},w_{K_-'})}
\prod_{K_+<K_+'}e^{-{k+2\over k}
D(\mu_{K_+},\mu_{K_+'};w_{K_+},w_{K_+'})
\ln E(w_{K_+},w_{K_+'})},\cr\eqinsert{\eqname{\iodd}}
}$$
where
$$\eqalignno{
D(\mu,\nu ; x,y)&=
\mu{\partial \over \partial x}
+\nu{\partial \over \partial y}
+\mu\nu{\partial \over \partial x}{\partial \over \partial y}.
&\eqname{\bibun}\cr}$$
For even genus, we obtain
$$\eqalignno{
\triangle^{(g:even)}_{[{\cal J},N',N;S,T,U]}(\tau)&=
\triangle^{(g:odd)}_{[{\cal J},N',N;S,T,U]}(\tau)\crr
&\qquad\times\exp\Bigl\lbrace-2\pi i\sum_{a=1}^g
\Bigl(2\sqrt2\alpha_0p^{[t_a]}_{\phi_0}
+\sqrt{k\over k+2}p^{[s_a,t_a]}_{\varphi}
\Bigr
)\int^Q_{P_0}d\omega^a\Bigr\rbrace,&\eqname{\trieven}\cr
{\cal E}^{(g:even)}(u,v,w)&={\cal E}^{(g:odd)}(u,v,w)
{\sigma(Q)^{c}\prod_{J}E(v_J,Q)^{{2\over p}-1}\over
\prod_{I}E(u_I,Q)
\prod_{K}E(w_K,Q)^{{2\over p'}-1}},&\eqname{\eeven}
\cr
I^{(g:even)}(\rho,\lambda,\mu,\nu)&=I^{(g:odd)}(\rho,\lambda,\mu,\nu)
\exp\Bigl\lbrace\sum_{K_+}\mu_{K_+}\partial_{w_{k_+}}
\ln E(w_{K_+},Q)
\Bigr\rbrace.&\eqname{\ieven}
\cr}$$
It is worthwhile to note the equation
$$\eqalignno{
&p^{[t_a]}_{\phi_0}\tau_{ab}p^{[t_b]}_{\phi_0}
+p^{[s_a,t_a,u_a]}_{\chi}\tau_{ab}p^{[s_b,t_b,u_b]}_{\chi}
-p^{[s_a,t_a]}_{\varphi}\tau_{ab}p^{[s_b,t_b]}_{\varphi}\cr
&=
{1\over 2\sqrt{kpp'}}(2pp'\Bigl[{t_a\over 2}\Bigr]+
(n_a'p-(-)^{t_a}n_ap')+k)\tau_{ab}
(2pp'\Bigl[{t_b\over 2}\Bigr]+(n_b'p-(-)^{t_b}n_bp')+k)\cr
&\qquad\qquad+2(k+2)\Bigl(\Bigl[{s_a\over 2}\Bigr]
-{u_a\over 2}-{J^{[s_a]}\over 2(k+2)}\Bigr)
\tau_{ab}\Bigl(\Bigl[{s_b\over 2}\Bigr]-
{u_b\over 2}-{J^{[s_b]}\over 2(k+2)}\Bigr)
\cr
&\qquad\qquad
-2k\Bigl({m^{[t_a]}\over 2k}+{u_a\over 2}\Bigr)\tau_{ab}
\Bigl({m^{[t_a]}\over 2k}+{u_a\over 2}\Bigr).
\cr}$$

\chapter{Topological Limit}
      We here discuss a topological limit, i.e. $l=0$, in  the
operator formalism given in the previous sections.
\par
      Firstof all, the central charge $c$ \centc\ vanishes,
in this limit, and
the integer $n$,
which lavel the primary field $\Psi_{J;n',n}$
is restricted to one due to \cosetprimb. Hence, only
the primary field of the type $\Psi_{J;n',1}$ survives.
Explicitly, such a field
is given by
$$\eqalignno{
\Phi_J\equiv&\Psi_{J;n',1}=e^{-{J\over \sqrt{k(k+2)}}(\varphi-\phi_0)}
&\eqname{\chiralp}\cr
}$$
with $J=n'-1$, $0\leq J\leq k$ . This field gives rise to
the chiral primary field in the N=2 super conformal theory.
The chiral primary ring is  as usual
$$\eqalignno{
\Phi_J\Phi_{J'}&=\Bigl\lbrace\matrix{\Phi_{J+J'} &for&0\leq J+J'\leq k\cr
                                             0  &\quad&otherwise.\cr}
&\eqname{\chiralf}\cr}$$
Due to  the same restriction $n=1$,
the $Q^+$ type BRST charge becomes just the screening charge
$Q^+_1=\oint dv S_+(v)$.
One should note that from \cscreenpm, the screening operator $S_+$
becomes  one of the super generator $G$ in the N=2 theory.
$$\eqalignno{
\lim_{l\rightarrow 0}S_+&=\Psi e^{\sqrt{k+2\over k}\phi_0}\equiv
\sqrt{k+2\over k}G.&\eqname{\superg}\cr
}$$
Hence, the BRST charge $Q^+_1
$ is identified with the one discussed by Eguchi and
Yang.\refmark{\eguyan}
\par
          The EM tensor $T$ is now BRST exact in the sense
$$\eqalignno{
T&=T_{Z_k}+{1\over 2}(\partial \phi_0)^2-{1\over 2}Q_{\phi_0}\partial^2
\phi_0
=\lbrace \bar G(z), Q^+_1\rbrace,
&\eqname{\topem} \cr}$$
where  $Q_{\phi_0}=-\sqrt{k\over k+2}$ and
$$\eqalignno{
\bar G&
=\sqrt{2k\over k+2}\Psi^{\dagger}e^{ -\sqrt{k+2\over k}\phi_0}.
&\eqname{\supergb}\cr}$$
\par
      The super partner of the chiral primary field $\Phi_J$ is given by
$$\eqalignno{
\Phi_J^{(1)}&=\oint dz\bar G(z)\Phi_J(w)
   =\sqrt{2\over k+2}J e^{-\sqrt{k+2\over k}({J\over k+2}-1)(\varphi
-\phi_0)}e^{-\chi}.&\eqname{\superp}\cr}$$
This satisfies the relation
$$\eqalignno{
\delta_B\Phi_J^{(1)}&=\partial \Phi_J.&\eqname{\brstmult}\cr}$$
Collecting the chiral and the
antichiral sector, one obtains a set of BRST invariant
observables:
$\Phi_J$, $\oint dw \Phi_J^{(1)}(w)$,
$ \oint d\bar w\bar\Phi_{\bar J}^{(1)}(\bar w)$ and
$ \int dw\wedge d\bar w \Phi_J^{(1)}(w)\bar\Phi_{\bar J}^{(1)}(\bar w)$.
\par
      The other BRST charge $Q_{J+1}$ is unchanged in the limit.
Hence the BRST cohomology groups as well as the trace formulae
for the topological minimal model
are obtained by setting  $l=0$ in \coscohom\ and \costrace.
\par
      It is instructive to make a connection to the manifestly
supersymmetric form of the model.\refmark{\yoshii,\lerda}
Let us define a pair of fields $\omega$ and $\sigma$ by the relations
$$\eqalignno{
\varphi-\phi_0&=\sqrt{k\over k+2}(\omega-\sigma),&\eqname{\osdefa}\cr
\varphi+\phi_0&=\sqrt{k+2\over k}(\omega+\sigma).
&\eqname{\osdefb}\cr
}$$
Then the bosonic pair $(\beta, \gamma)$\foot{
Do not confuse this with $(\beta, \gamma)$ in $\S$2 and $\S$3.}
 and the fermionic pair
$(b,c)$ defined by
$$\eqalignno{
\beta&=e^{-\omega}\partial e^{\chi},\qquad
\gamma=e^{\omega}e^{-\chi},&\eqname{\bcdefa}\cr
b&=e^{-\sigma},\qquad c=e^{\sigma}&\eqname{\bcdefb}\cr
}$$
give rise to a supersymmetric multiplet of
spin $\lambda={k+1\over k+2}$ conjugate first order
systems.\refmark{\fms}
In terms of these fields,
 the EM tensor \topem\ and the super generators $G$ and $\bar G$
are reexpressed as
$$\eqalignno{
T&=-\lambda b\partial c+(1-\lambda)(\partial b)c-
\lambda \beta\partial\gamma+(1-\lambda)(\partial\beta)c,&\eqname{\stopem}\cr
G&\sim \beta c,&\eqname{\ssuperg}\cr
\bar G&\sim \lambda(\partial \gamma)b-(1-\lambda)
\gamma\partial b.&\eqname{\ssupergb}\cr
}$$
\par
    Let us next consider
the calculation of correlation functions.
In the operator formalism discussed in the previous sections,
we have to consider the following changes.
First, the screening charge of the type $\oint S_+$ can no longer
be inserted  into any  correlators. Otherwise it causes vanishing results.
Therefore, in the screened three string vertex and
the screened $g$-loop operator, we set $r_+=0$.
Secondly, we have to consider a coupling with the super partner
$\Phi^{(1)}_J$.
However, for  the correlator with any nonzero number of $\Phi^{(1)}_J$,
it is  not possible to  screen the whole background charges by making use of
the screening operators $S$ and $S_-$ only.
Hence one has to add  another set of screening operators.
A candidate is  the one  appearing in the $Z_k$ parafermion theory:
\refmark{\distqiu,\frau,\jns}
$$\eqalignno{
V_+&=e^{-\chi}, \qquad
V_-=e^{\sqrt{k(k+2)}\varphi}e^{-(k+1)\chi}
\cr}$$
The charge $ Q_V=\oint  V_+$ has been used in section 3 to
subtract redundant zero modes of $\eta$. There, the Fock space
of the $Z_k$ parafermion sector is projected onto the kernel of
$ Q_V$. Therefore only $V_-$ can be used in the correlators.
\par
      By adding
the screening charge $\oint V_-$, $r_V$ times,
to the screened $g$-loop operator
with properly chosen integration contours,
the selection rules for the
correlator
$<{\cal T}^{g(r,0,r_-,r_V)}\vert$\break
$\prod_i^N\Phi_{J_i}(z_i)
\prod_j^M\oint dy_j\Phi^{(1)}_{J_j}(y_j)
\vert 0>$
are given by
$$\eqalignno{
&\sum_{i}^N{J_i\over \sqrt{k(k+2)}}+
\sum_{j}^M\sqrt{{k+2\over k}}(-1+{J_j\over k+2})
-{2r_-\over \sqrt{k(k+2)}}
+\sqrt{{k\over k+2}}(g-1)=0,&\eqname{\toposela}
\cr
&
\sum_{i}^N{J_i\over \sqrt{k(k+2)}}+
\sum_{j}^M\sqrt{k+2\over k}(-1+{J_j\over k+2})
+\sqrt{k\over k+2}r-\sqrt{k+2\over k}r_-
\cr
&\qquad\qquad\qquad\qquad\qquad\qquad-\sqrt{k(k+2)}r_V
+\sqrt{{k\over k+2}}(g-1)=0,&\eqname{\toposelb}
\cr
&r- r_--M-(k+1)r_V=0.&\eqname{\toposelc}
\cr}$$
Solving these equations, we obtain the equivalent conditions:
$r_V=M$, $r-r_-=(k+2)r_V$ and \toposela.
\par
      Note that if one sets $r_-=0$, the condition \toposela\  coincides
with the one obtained by Li.\refmark{\keli}, whereas
the scheme using nonzero $r_-$
has been  discussed by Kawai et.al.\refmark{\kawai}.
\par
      Let us take
the minimal choice of screening operators,  $r_-=0,
r_V=M$ and $r=(k+2)M$, following Li.
Making use of the operator formalism with $l=0$
discussed in the previous sections,
one can now in principle
calculate any correlators in the topological minimal
model.\par
     It is, however, a formidable task to carrying out the whole
integrations associated with $\Phi^{(1)}_J$ and screening charges.
The root of this troublesome feature is however simple.
Namely, it lies in the extra $\chi$-charge in $\Phi^{(1)}_J$, which
requires nonzero $r$ and $r_V$.
Hence, if  a picture of $\Phi^{(1)}_J$ can be changed to
the one with  $\chi$-charge zero, it becomes possible to
make a calculation of arbitrary genus correlator without any
screening charges.
\par
       One possible picture-changing operation is as follows.
$$\eqalignno{
\Phi^{(1)'}_J(w)&=\lbrace Q^+_1, \tilde I(w)\Phi^{(1)}_J(w)\rbrace
                ={J\over J-k-2}\partial\Bigl(
                   e^{-\sqrt{k+2\over k}(-1+{J\over k+2})(\varphi-\phi_0)(w)}
                    \Bigr),&\eqname{\pictchange}\cr
}$$
where $\tilde I$ is an identity operator given by
$$\eqalignno{
\tilde I(w)&=e^{\sqrt{k+2\over k}(\varphi-\phi_0)(w)}.&\eqname{\tildeid}\cr
}$$
The operator $\tilde I$ has
just the same amount of
$\varphi$- and $\phi_0$-charges as $Q^+_1$ with the opposite sign
so that the picture-changed field $\Phi^{(1)'}_J(w)$ has
the same $\varphi$- and $\phi_0$-charges
as $\Phi_J^{(1)}$.\par
      Note that, in \pictchange,
the picture-changing operator is identified with
$$\eqalignno{
G(z)\tilde I(w)&\sim \partial e^{\chi(w)}.&\eqname{\picchop}\cr
}$$
A similar picture changing operator has been discussed by
Distler\refmark{\distler} in the context of two dimensional topological
gravity.\foot{Naively, the fields in
Distler's 2D topological gravity are
obtained by  \bcdefa\ and \bcdefb with
the level number $k$ being analytically continued to -3.}
\par
       Since the $\chi$-charge zero picture operator
$\Phi^{(1)'}_J$ is a total derivative, a naive calculation using
$\Phi^{(1)'}_J$ gives rise to a vanishing result.
However, as discussed by Distler, an insertion of the picture
changing operator causes a singularity so that one has to
consider a certain regularization of the integral.
If a proper regularization
scheme is found, one could expect
that the integral of the above total derivative term gives
rise to a representative of a certain cohomology class
associated with the topological minimal model.
To find
such a regularization scheme is an open problem.
\par
      The author would like to thank Prof.M.Kato and Prof.A.Nakayashiki
for discussions.
He is also grateful to Prof.N.Nakanishi and
Prof.I.Ojima for reading the manuscript.

\Appendix{A}
      We here give a derivation of the homology cycles
${\cal C}, {\cal C}_{+}, {\cal C}_-$ in \sccsv.
Let us consider
a form dual to ${\cal C}\wedge{\cal C}_{+}\wedge{\cal C}_-$:
$$\eqalignno{
\omega&=\prod_i^rdu_i\prod_j^{r_+}dv_j\prod_k^{r_-}dw_k\cr
&\qquad\times _3<J_3;n_3',n_3|<V_{123}>
\prod_{i=1}^r S^{(1)}(u_i)\prod_{j=1}^{r_+}
S_+^{(1)}(v_j)\prod_{k=1}^{r_-}S_-^{(1)}(w_k)
|J_1;n_1',n_1>_1|J_2;n_2',n_2>_2
\cr
&=
\prod_i^rdu_i\prod_j^{r_+}dv_j\prod_k^{r_-}dw_k
f(u_i,v_j,w_k)
\prod_{i,k}(u_i-w_k)^{-2}
\prod_{j,k}(v_j-w_k)^{-2}\cr
&\qquad\qquad\times\prod_{i=1}^ru_i^{-{J_1\over k+2}-1}
(1-u_i)^{-{J_2\over k+2}-1}
\prod_{i<i'}^r(u_i-u_{i'})^{2\over k+2} \cr
&\qquad\qquad\times\prod_{j=1}^{r_+}v_j^{{1-n_1\over p}-1}
(1-v_j)^{{1-n_2\over p}-1}
\prod_{j<j'}^r(v_j-v_{j'})^{{2\over p}}\cr
&\qquad\qquad\times\prod_{k=1}^{r_-}w_k^{-{1-n_1'\over p'}-1}
(1-w_k)^{-{1-n_2'\over p'}-1}
\prod_{k<k'}^r(w_k-w_{k'})^{-{2\over p'}-2}, &\eqname{\formo}
\cr}$$
with $r=r_--r_+$, $2r_+=n_1+n_2-n_3-1$ and $2r_-=n_1'+n_2'-n_3'-1$.
In \formo\ ,
the source points are chosen
as $z_1=0, z_2=1$ and $z_3=\infty$, and
$f$ is a holomorphic, symmetric and single-valued function.
A relevant homology group dual to the cohomology of the forms in \formo\
 is
defined on a pair of circles, one  of
which winds the source points
0 (=$z_1$) and the other does 1 (=$z_2$) once,
and contact each other at one  point.
Defining the r-chains $C^{(\pm)}_j$
 and r-1-chains $S^{(\pm)}_j, j=0,1,\cdots, r_{(\pm)}$
according to Ref.[\fesi],
we find that the boundaries of the $j$th $r_{(\pm)}$-chains $C^{(\pm)}_j$
are given by
$$\eqalignno{
\partial C_j&=(q^{r-J_2-j-1}-1)S_{j}-(q^{J_1-j+1}-1)S_{j-1}
&\eqname{\boundarya}
\cr
\partial C^+_j&=(q_+^{r_+-n_2-j}-1)S^+_{j}-(q_+^{n_1-j}-1)S^+_{j-1}
&\eqname{\boundaryb}
\cr
\partial C^-_j&=(q_-^{r_--n_2'-j}-1)S^-_{j}-(q_-^{n_1'-j}-1)S^-_{j-1}
&\eqname{\boundaryc}
\cr
}$$
The desired cycles are thus obtained by
$$\eqalignno{
{\cal C}&=\sum_{j=0}^r {\cal A}_j C_j,\qquad
{\cal C}_+=\sum_{j=0}^{r_+} {\cal A}^+_j C^+_j,\qquad
{\cal C}_-=\sum_{j=0}^{r_-} {\cal A}^-_j C^-_j,&\eqname{\cycles}\cr
}$$
with ${\cal A}_0={\cal A}^+_0={\cal A}^-_0=1$ and
$$\eqalignno{
 {\cal A}_j&=\prod_{k=1}^j{(q^{r-J_2-k}-1)\over
(q^{J_1'-k+1}-1)},&\eqname{\cnocoeffa}
\cr
{\cal A}^+_j&=\prod_{k=1}^j{(q_+^{r_+-n_2-k+1}-1)\over
(q^{n_1-k}-1)},&\eqname{\cnocoeffb}
\cr
{\cal A}^-_j&=\prod_{k=1}^j{(q^{r_-n_2'-k+1}-1)\over
(q^{n_1'-k}-1)}.&\eqname{\cnocoeffc}
\cr
}$$
One can easily check $\partial {\cal C}=0, \partial {\cal C}_+=0$
and $\partial {\cal C}_-=0$ from \boundarya $\sim$ \boundaryc.
\par

\Appendix{B}
      We here give a proof of the BRST relations in  \brela\ and \brelb.
\par
      The line of proof is the same as that in  the $SU(2)_k$ WZW model.
\refmark{\konnow}
Using  the overlapping condition, we obtain
$$\eqalignno{
&Q_{J+1}^{(3)}<V_{123}^{(r,r_+,r_-)}>\cr
&=-<V_{123}^{(r,r_+,r_-)}>
{\cal C}^J_1P_JP_{Jr}Q_{J+1}^{(1)}-({\cal C}^A_3P_A)^{-1}
<V_{123}^{(r+J-A,r_+,r_-)}> Q_{A+1}^{(2)},
&\eqname{\appe}\cr}$$
where the phase factors
${\cal C}^J_a, a=1,3, P_J$ and $  P_{Jr}$ , arise
from the overlapping condition, from
the commutation
among $J+1$ $S$'s and from the one between
$J+1$ and $r$ $S$'s, respectively;
$$\eqalignno{
{\cal C}^J_a&=
e^{i\pi(J+1)(-\sqrt{k\over k+2}\alpha_{\varphi,0}^{(a)}
+\alpha_{\chi,0}^{(a)})},&\eqname{\phasec}\cr
P_J&=e^{i\pi J(J+1)(-{k\over k+2}+1)/2},&\eqname{\phasep}\cr
P_{Jr}&=e^{i\pi r(J+1)(-{k\over k+2}+1)}.&\eqname{\phaseb}\cr
}$$
Similarly, we obtain for the BRST charge $Q^+_n$
$$\eqalignno{
&Q_{n}^{+(3)}<V_{123}^{(r,r_+,r_-)}>\cr
&=-<V_{123}^{(r,r_+,r_-)}>
{\cal C}^{+n}_1P_nP_{nr_+}Q_{n}^{+(1)}-({\cal C}^{+m}_3P_m)^{-1}
<V_{123}^{(r,r_++n-m,r_-)}> Q_{m}^{+(2)},
&\eqname{\appee}\cr}$$
where
$$\eqalignno{
{\cal C}^{+n}_a&=
e^{i\pi n(-\sqrt{k+2\over k}\alpha_{\varphi,0}^{(a)}+\alpha_{\chi,0}^{(a)}
+\sqrt2\alpha_+\alpha_{\phi_0,0}^{(a)})
},&\eqname{\pasc}\cr
P_n&=e^{i\pi n(n-1)(\alpha_+^2-{1\over k})},&\eqname{\pasb}\cr
P_{nr_+}&=e^{i\pi nr_+(-{k+2\over k}+1+2\alpha_+^2)},&\eqname{\pasa}\cr
}$$
Replacing the momentum operator
$\alpha^{(3)}_{\varphi,0},\alpha^{(3)}_{\chi,0}$
and $\alpha^{(3)}_{\phi_0,0}$ with their eigenvalues\break
${1\over 2}\sqrt{k\over k+2}({k+2\over k}m_{n',n}
-J),\quad -{1\over 2}(J-m_{n',n})$
and $\sqrt2\alpha_{n',n}$, and using  the charge conservation law \chargecon,
one gets the BRST relations \brela\ and \brelb.
\refout
\endpage
\end
\bye